%% file: Paper18.tex
\documentclass[submission,copyright,creativecommons]{eptcs}
\providecommand{\event}{TTC 2011}
\usepackage{latexsym}
\usepackage{amssymb}
\usepackage{extarrows}
\usepackage{epsfig}
\usepackage{graphicx}
\usepackage{amsmath}
\usepackage{fancyhdr}
\usepackage{cite}
\usepackage{multirow}
\usepackage{listings}
\usepackage{booktabs}
\usepackage{amsfonts}
\usepackage{times}
\usepackage{fleqn}
\usepackage{macros}
\usepackage{algorithm}
\usepackage{algorithmic}
\usepackage{hyperref}
\usepackage{breakurl}

\newcommand{\hide}[1]{}

\title{\centering Saying Hello World with QVTR-XSLT \\ A Solution to the TTC 2011 Instructive
Case}
\author{Dan Li \thanks{On leave from Guizhou Academy of Sciences, Guizhou, China}, Xiaoshan Li
\institute{Faculty of Science and Technology, University of Macau, China}
\email{lidan@iist.unu.edu, xsl@umac.mo}
\and
Volker Stolz
\institute{Department of Informatics, University of Oslo, Norway \& UNU-IIST, Macau, China}
\email{stolz@ifi.uio.no}
}

\def\titlerunning{Saying Hello World with QVTR-XSLT}
\def\authorrunning{Dan Li, Xiaoshan Li \& Volker Stolz}

\begin{document}
\maketitle

\section{Introduction \label{se:introduction}}

In this short paper we present our solution for the Hello World case study \cite{helloworldcase} of the Transformation Tool Contest (TTC) 2011 using the \href{http://rcos.iist.unu.edu/qvttoxslt/}{QVTR-XSLT tool} \cite{LiQVTUMLFM2011}.
The tool supports editing and execution of the graphical notation of QVT Relations language \cite{OMG_QVT20}.

The case study consists of a set of simple transformation tasks which covers the basic functions required for a transformation language, such as creating, reading/querying, updating and deleting of model elements.
We design a transformation for each of the tasks.
The \verb"SHARE" demo related to the paper can be found at \cite{QVTRXSLTsolutions}.

We begin by giving a brief introduction of the QVTR-XSLT tool in Section~\ref{se:qvtrtool}.
Section~\ref{se:casestudy} provides  the solutions for the tasks of the case study.
We discuss the conclusion in Section~\ref{se:conclusion}.
Details of the transformation definitions are presented in the appendices.

\input{sec_QVTXSLT}

\section{Solution \label{se:casestudy}}


As the first step for transformation design, we define all metamodels described in the case specification using the graphical editor of the QVTR-XSLT tool. 
Simple UML class diagrams are used to specify metamodels in the tool.
The source models provided by the case study conform  the metamodels.
In addition, we design a simple HTML metamodel as the target metamodel for the model-to-text transformation.
We also have a \emph{Result} metamodel to store the \emph{results} of querying matched model elements, as well as the descriptions of the queries.
Appendix~\ref{ap:metamodels} (Fig.~\ref{fig:HelloWorldmeta}--\ref{fig:MoreEvolvedGraph}) shows all the metamodels.


\subsection{Hello world}

This task consists of three subtasks: two constant transformations and a  model-to-text transformation.
We complete each subtask using a transformation of a single relation, as shown in Appendix~\ref{ap:helloworld}.
There must be a source model for a transformation, even we may not actually need any information of the model.
For the convenience we take the \emph{SimpleGraph} metamodel as the source metamodel for the constant transformations, because the case study has already provided a model of the metamode.
We use the \emph{HtmlMetaModel} as the target metamodel of the model-to-text transformation, so its output is a web page, which can be easily displayed in a browser.

\subsection{Count matches with certain properties}

The task focuses on querying a simple graph model to find elements with certain properties.
We design four \emph{Queries} (Fig.~\ref{fig:GetNodesNumber}--\ref{fig:GetDanglingEdges}) using the graphical notation for counting the number of nodes, isolated nodes, looping edges and dangling edges.
In all these queries, OCL function \emph{size()} is used to count the number of elements satisfying the conditions, and a predefined variable \emph{result} returns the results of the queries.
Moreover, we define a function \emph{GetAllCircleNodes} (Fig.~\ref{fig:GetAllCircleNodes}) for counting the number of matches of a circle consisting of three nodes.
The function is directly written in XSLT, the back-end language of our tool, and calls within it another XSLT function \emph{GetCircleNodes} and a query \emph{GetLinkedNodes} (Fig.~\ref{fig:GetLinkedNodes}), which returns all target nodes of a given source node.
In actually, the two functions offer the capacity to calculate general \emph{k}-circles, and the \emph{k} is given as the last parameter of function \emph{GetCircleNodes}.
It shows how XSLT code can be seamlessly integrated with QVTR to provide more powerful features.

The transformation starts from the relation \emph{GraphToResult} (Fig.~\ref{fig:GraphToResult}), in where queries and functions are invoked one by one, and the count results and their explanations are sent to relation \emph{ShowIntResult} for adding to the result model.

\hide{
\begin{table}[!h]
 \centering \caption{Transformations and experiment results }
\label{tab:result}
 \begin{tabular}{lllll}
  \toprule
 Task & \ \   Number of   & \ \  Lines & \ \  Input model & \ \  Exec  \\
     &   \ \     relations   & \  \ of & \ \ (.xmi)  & \ \  time  \\
     &   \ \      /queries/functions  & \ \  XSLT &  & \ \  (ms) \\
  \midrule
Hello world (constant)& \ \  1 & \ \  47 & \ \  Graph1 & \ \  $<1$ \\

Hello world (extended constant) & \ \  1 & \ \  52 & \ \  Graph1 & \ \  $<1$ \\

Hello world (model-to-text)  & \ \  1 & \ \  60 & \ \  Greetingext & \ \  $<1$ \\

Count matches & \ \  2/5/2 & \ \  172 & \ \  Graph1 & \ \  13 \\

Reverse edges & \ \   3 & \ \  81 & \ \  Graph1 & \ \   $<1$ \\

Simple migration & \ \   3 & \ \  82 & \ \  Graph1 & \ \  $<1$ \\

Topology-changing migration & \ \   2/1 & \ \  73 & \ \  Graph1 & \ \  $<1$ \\

Delete node & \ \   2 & \ \  81 & \ \  Graph1 & \ \  $<1$ \\

Insert transitive edges & \ \   5/2 & \ \  139 & \ \  Graph1 & \ \  15 \\

  \bottomrule
 \end{tabular}
\end{table}
}

\subsection{Reverse edges}

The transformation takes the \emph{SimpleGraph} as both the source and target metamodels. 
We design three top-level relations for copying graphs, copying nodes and copying edges while exchanging the source and target nodes, as shown in Appendix \ref{ap:reverseedges}.
This transformation can properly handle dangling edges.

\subsection{Simple migration}

Using the \emph{EvolveGraph} as the target metamodel, the transformation is also designed as three top-level relations for migrating graphs, nodes and edges respectively (see Appendix \ref{ap:simplemigration}).

\subsection{Topology-changing migration }

With the \emph{MoreEvolveGraph} as the target metamodel, the transformation only has two relations to migrate graphs and nodes, and a query is used to get the destination nodes of a node (see Appendix \ref{ap:topology}).

\subsection{Delete node with specific name and its incident edges}

Different from all above transformations, the task is completed by an \emph{in-place} transformation, where two relations are used to mark the node with name "n1", along with the incident edges of the node, as \emph{remove} in the \emph{xmiDiffOp} property (see Appendix \ref{ap:deletenode}).
For an in-place transformation, the source and target models are the same during the execution, and the model elements can be added, deleted and updated.

When an in-place transformation is executed, the modifications to the model will be collected into a \emph{difference model} as a set of adding, deleting, or altering operations.
Then the source model is modified according to the records of the \emph{difference model} to get the result model.
For example, the node and the edges will be deleted in this task.
This process is user transparent and runs automatically.

\subsection{Insert transitive edges}

Similar to the other transformations, we use three relations to copy graphs, nodes, and edges from the source model to the target model. 
Furthermore, with the help of two queries that obtain a node's destination nodes (except the node itself) and the edges from one node to another node, the relation \emph{LookTransitive} calculates the transitive closure, and invokes relation \emph{InsertEdge} to insert an edge between the two indirectly related nodes.
Appendix~\ref{ap:insertedges} depicts the definitions of the transformation.

\section{Experiments and Conclusion \label{se:conclusion}}

\begin{table}[!h]
 \centering \caption{Transformations and experiment results }
\label{tab:result}
 \begin{tabular}{lllll}
  \toprule
 Task & \ \   Number of   & \ \  Lines & \ \  Input model & \ \  Exec  \\
     &   \ \     relations   & \  \ of & \ \ (.xmi)  & \ \  time  \\
     &   \ \      /queries/functions  & \ \  XSLT &  & \ \  (ms) \\
  \midrule
Hello world (constant)& \ \  1 & \ \  47 & \ \  Graph1 & \ \  $<1$ \\

Hello world (extended constant) & \ \  1 & \ \  52 & \ \  Graph1 & \ \  $<1$ \\

Hello world (model-to-text)  & \ \  1 & \ \  60 & \ \  Greetingext & \ \  $<1$ \\

Count matches & \ \  2/5/2 & \ \  172 & \ \  Graph1 & \ \  13 \\

Reverse edges & \ \   3 & \ \  81 & \ \  Graph1 & \ \   $<1$ \\

Simple migration & \ \   3 & \ \  82 & \ \  Graph1 & \ \  $<1$ \\

Topology-changing migration & \ \   2/1 & \ \  73 & \ \  Graph1 & \ \  $<1$ \\

Delete node & \ \   2 & \ \  81 & \ \  Graph1 & \ \  $<1$ \\

Insert transitive edges & \ \   5/2 & \ \  139 & \ \  Graph1 & \ \  15 \\

  \bottomrule
 \end{tabular}
\end{table}

We have solved all mandatory and optional tasks of the case study. 
We generate XSLT stylesheets for all the transformations through our code generator.
Table~\ref{tab:result} shows the number of relations/queries/functions and lines of XSLT code for each transformation.
There are only about 30 lines of XSLT code that are directly hand-written and embed into the transformation counting matches, and all others are generated from the graphical notation of QVTR.

Using our transformation runner, the transformations are executed in a laptop of Intel  2.13 GHz M330 CPU, 3 GB memory, and running Windows 7 home version.
The results are also shown in Table~\ref{tab:result}.
The execution time includes the time for loading and saving model files from/to disk.


\paragraph{Conclusion \\}

We presented transformations for the Hello World case study of TCC 2011 to show how basic model transformation problems can be solved with the QVTR-XSLT tool.
These transformations are designed using the standard graphical notation and OCL expressions of QVT Relations in a straightforward, concise and intuitive way.
We hope the case study will help to demonstrate that the language and the tool can be efficiently applied to model transformations in practice.

{\small
\noindent{\bf \\ Acknowledgements } Partially supported by the ARV and GAVES grants of the Macau Science and Technology Development Fund, and the Guizhou International Scientific Cooperation Project G[2011]7023 and GY[2010]3033.
}

\bibliographystyle{eptcs}
{\small \bibliography{bibtex/managed,bibtex/rcos}}

\newpage

\appendix
\input{TTC2011_Appendix_qvtrintro.tex}

\clearpage
\input{TTC2011_Appendix_hello.tex}

\end{document}

%% file: sec_QVTXSLT.tex
\section{The QVTR-XSLT tool \label{se:qvtrtool}}

Model transformation is the core technology for model-driven development, and is used in software model refinement, evolution, refactoring and code generation.
To address the need for a common transformation language, the Object Management Group (OMG) proposed the Query/View/Transformation language (QVT) \cite{OMG_QVT20} standard.
QVT has a hybrid declarative/imperative nature.
In its declarative language, called QVT Relations (QVTR), a transformation is defined as a set of \emph{relations} (rules) between source and target models, each conforming to their respective metamodels.
Transformations are driven by a single, designated top-level relation.

QVTR combines a textual and a graphical notation.
In graphical syntax, a relation specifies how two object diagrams, called \emph{domain patterns}, relate to each other.
That is, the \emph{structural} matching of elements in the source- and target model is done diagrammatically.
Moreover, QVTR employs a textual language based on essential OCL \cite{OMG_OCL20} to define additional (non-structural) constrains in relations.
The graphical notation of QVTR provides a concise, intuitive and yet powerful way to specify transformations. However, currently there are very few tools supporting QVTR, and even fewer for its graphical notation.
\hide{
QVTR-XSLT is a tool supporting a \underline{slightly extended}\footnote{Hm, so you have some extensions, but you do not support ``full'' QVTR. This makes it very easy to attack the paper. What are the extensions?} version of the graphical notation of QVT Relations. It consists of two parts:
}

\begin{figure}[!h]
\begin{minipage}[t]{0.5\linewidth}
\vspace{0pt} \centering
\includegraphics[width=1.0\linewidth]{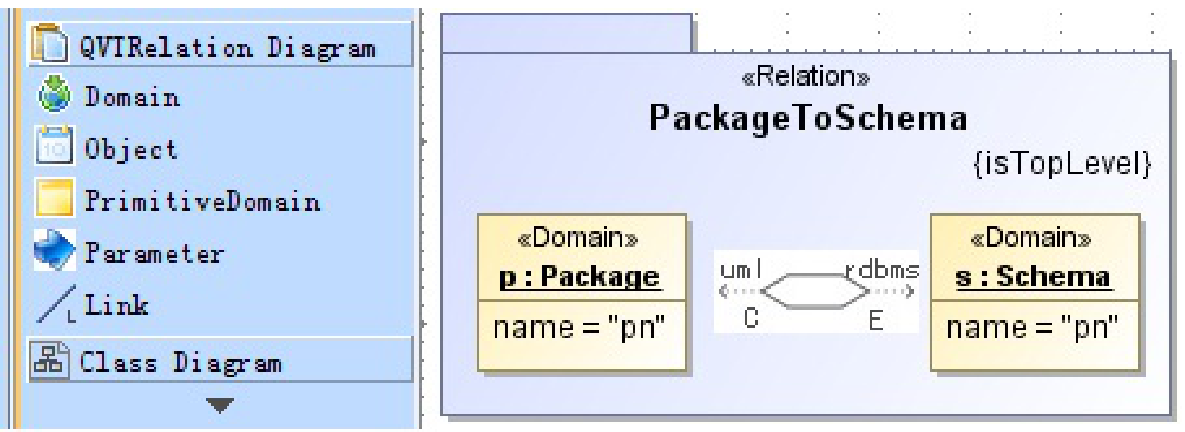}
\end{minipage}%
\begin{minipage}[t]{0.5\linewidth}
\vspace{0pt} \centering
\includegraphics[width=0.9\linewidth]{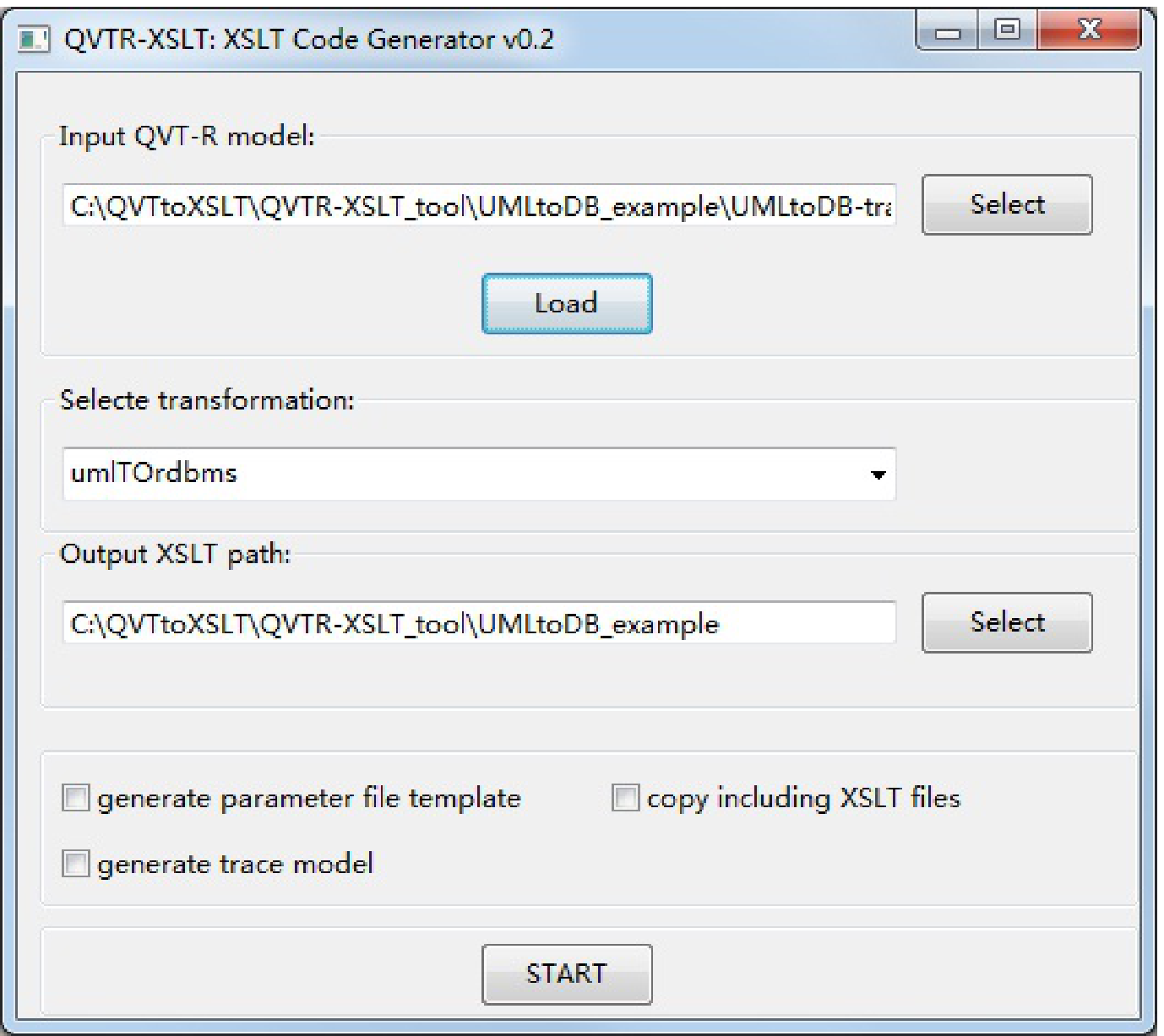}
\end{minipage}%
\\[+2pt]
\begin{minipage}[c]{0.5\linewidth}
\caption{Toolbar of QVTR graphical editor}
\label{fig:QVTRToolbar}
\end{minipage}%
\begin{minipage}[c]{0.5\linewidth}
   \caption{XSLT code generator}
\label{fig:codegenerator}
\end{minipage}
\end{figure}

QVTR-XSLT supports the graphical notation of QVT Relations, and an execution engine for a subset of QVTR by means of XSLT programs.
It consists of two parts:

\begin{itemize}
  \item \textbf{Graphical Editor}: Building on top of \emph{MagicDraw UML} \cite{MagicDraw}, the editor has a graphical interface for defining metamodels as simple class diagrams, specifying QVTR relations and queries in graphical notation, validating the design, and saving the transformations as an XML file. The toolbar of the graphical editor is showed in Fig.~\ref{fig:QVTRToolbar}.
  \item \textbf{Code generator}: It reads in the XML file, and generates an XSLT stylesheet for a transformation. Fig.~\ref{fig:codegenerator} illustrates the interface of the code generator.
\end{itemize}

\vspace*{-1.0\baselineskip}

The outputs of the code generator are pure XSLT programs which can be directly executed under any XSLT processor on any platform.
Additionally, we have also developed a transformation runner, in the form of a Java program invoking the Saxon 9 XSLT processor, to facilitate the execution of generated XSLT stylesheets.

The QVTR-XSLT tool supports transformation parameters, transformation inheritance through rule overriding, and multiple input and output models.
Furthermore, \emph{in-place} transformations are defined as modifications (insert, remove, replace) of the existing model elements.
QVTR-XSLT-based transformations are used in the \href{http://rcos.iist.unu.edu}{rCOS Modeler} for use case-driven development of component- and object systems.


%% file: TTC2011_Appendix_qvtrintro.tex
\section{A Brief Introduction to QVT Relations \label{ap:qvtintro}}

QVT Relations (QVTR) is a declarative model transformation language proposed by the OMG as part of the MOF Query/View/Transformations (QVT) standard \cite{OMG_QVT20}.
QVTR specifies a \emph{transformation} as a set of \emph{relations} between
source and target metamodels.
A metamodel is defined in our tool as a simple class diagram.
In addition, a transformation may own some \emph{functions}, which are side-effect-free
operations, and \emph{queries} used to retrieve information from the models.

In the graphical notation, a \emph{relation} defines how two object diagrams, called \emph{domain
patterns}, relate to each other.
The object with  tag \emph{$\ll$domain$\gg$} is the \emph{root} of a domain pattern, and it also serves as a parameter of the relation.
In general, we assume the left domain pattern is the source domain, and the right the target domain.
An \emph{object} or a property of an object could be given a name that is taken as a \emph{variable}.
If the object is in the source domain pattern, then the object or the value of the property is bound to the variable.
Otherwise the object in target domain pattern means assigning the value of the variable to the object or property.
Note that a property variable in the diagrams may contain additional quote-characters that are an artefact of the visualization, and not string delimiters.

When a relation is executed, the source domain pattern is searched in the source model by way of
\emph{pattern matching} which starts from the domain root.
When a match is found, all variables defined in source domain pattern are bound to values.
The target domain pattern acts as a template to create corresponding objects and links in the target model using the values of the variables in the pattern.

A relation may define a pair of optional \emph{when}- and \emph{where}-clauses which consist of a set of OCL expressions.
The \emph{when}-clause indicates additional matching conditions for the relation.
And new variables can be defined in the \emph{where}-clause.
Other relations could be invoked in the \emph{where}-clause and variables can be passed as arguments.
A relation may also have \emph{primitive domains} in order to pass additional parameters between the relations.
Furthermore, a relation is either designed as a \emph{top-level} relation, or
a \emph{non-top-level} relation.
A \emph{top-level} relation is invoked from the transformation framework, and \emph{non-top-level} relations are invoked by other relations.

%% file: TTC2011_Appendix_hello.tex
\section{Metamodel definitions \label{ap:metamodels}}

\vspace*{-1.5\baselineskip}
\begin{figure}[!h]
\begin{minipage}[t]{0.2\linewidth}
\vspace{0pt} \centering
\includegraphics[width=1.0\linewidth]{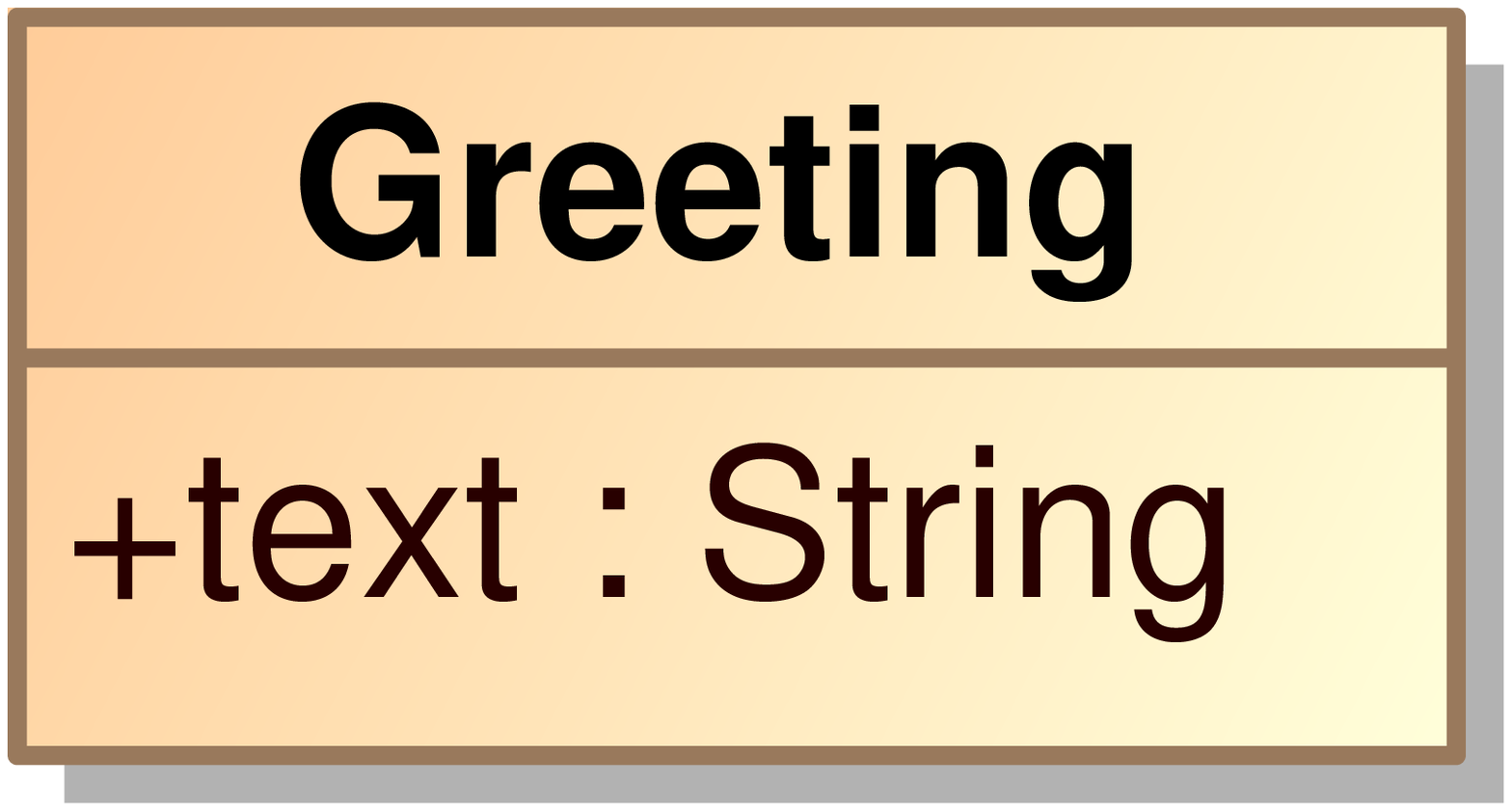}
\end{minipage}%
\begin{minipage}[t]{0.4\linewidth}
\vspace{0pt} \centering
\includegraphics[width=.9\linewidth]{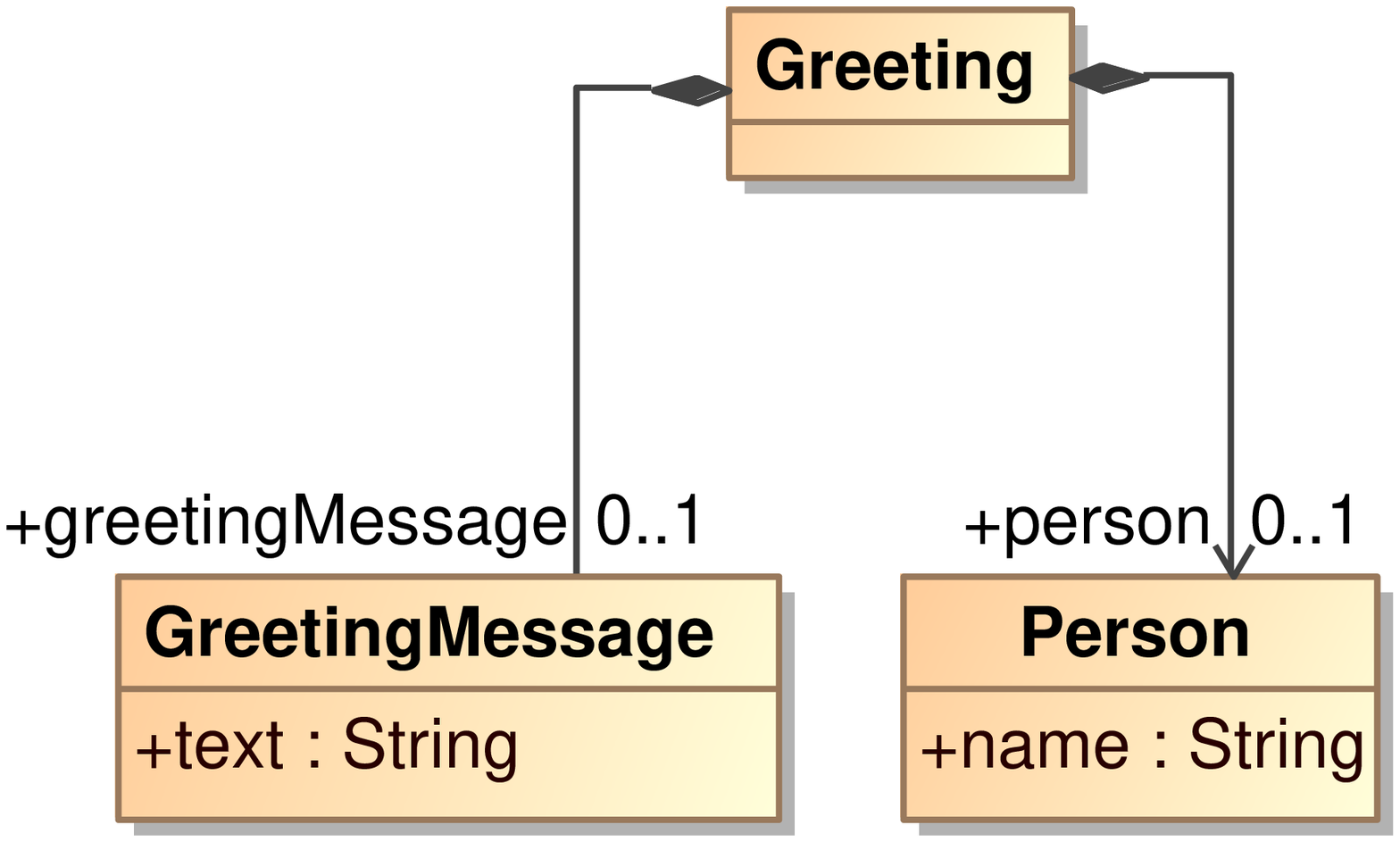}
\end{minipage}%
\begin{minipage}[t]{0.4\linewidth}
\vspace{0pt} \centering
\includegraphics[width=0.8\linewidth]{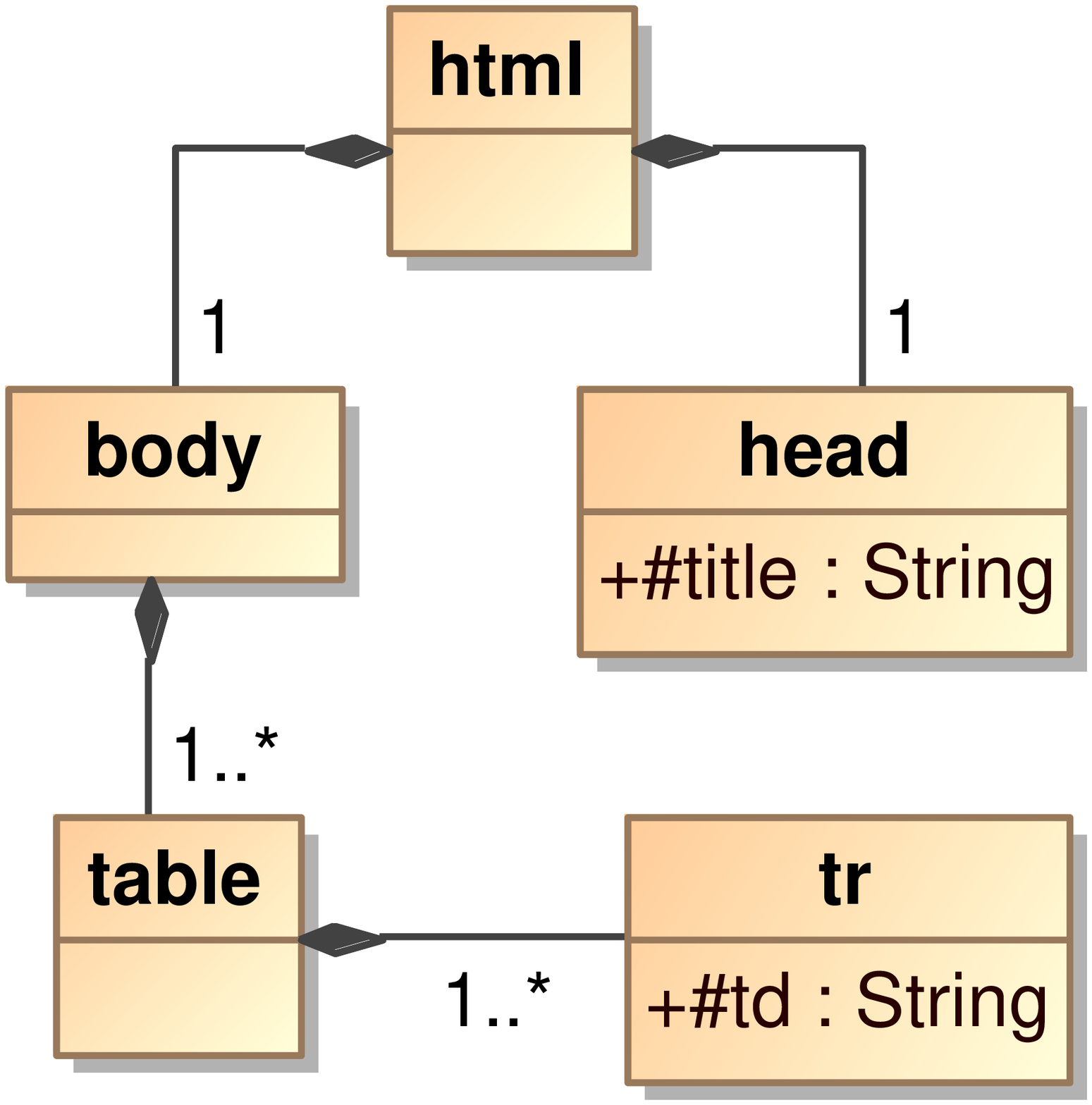}
\end{minipage}%
\\[+2pt]
\begin{minipage}[c]{0.3\linewidth}
\caption{HelloWorld}
\label{fig:HelloWorldmeta}
\end{minipage}%
\begin{minipage}[c]{0.3\linewidth}
   \caption{HelloWorldExt}
\label{fig:HelloWorldExtmeta}
\end{minipage}
\begin{minipage}[c]{0.4\linewidth}
   \caption{HtmlMetaModel}
\label{fig:HtmlMetaModel}
\end{minipage}
\end{figure}

\vspace*{-1.5\baselineskip}

\begin{figure}[!h]
\begin{minipage}[t]{0.5\linewidth}
\vspace{0pt} \centering
\includegraphics[width=.9\linewidth]{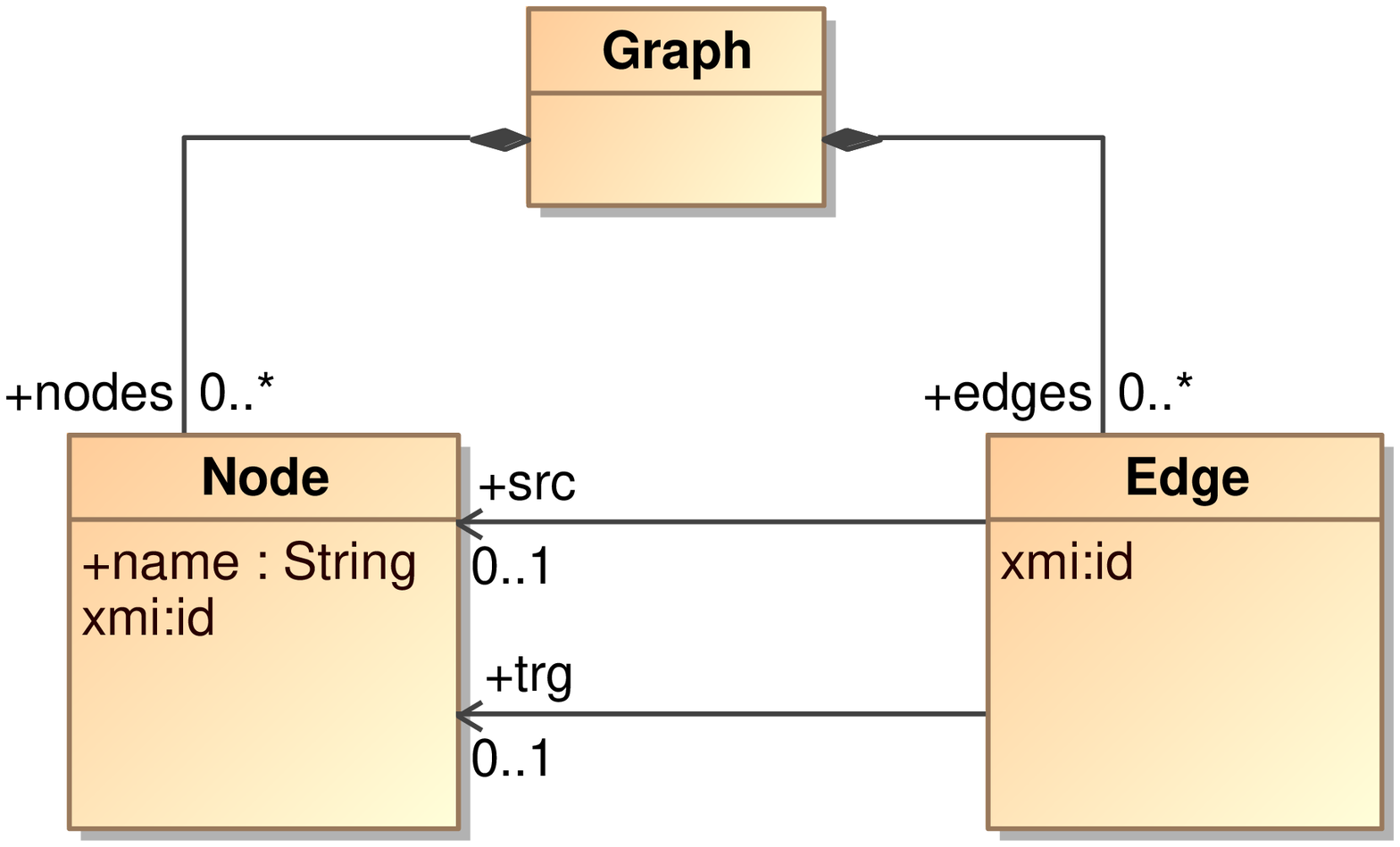}
\end{minipage}%
\begin{minipage}[t]{0.5\linewidth}
\vspace{0pt} \centering
\includegraphics[width=.9\linewidth]{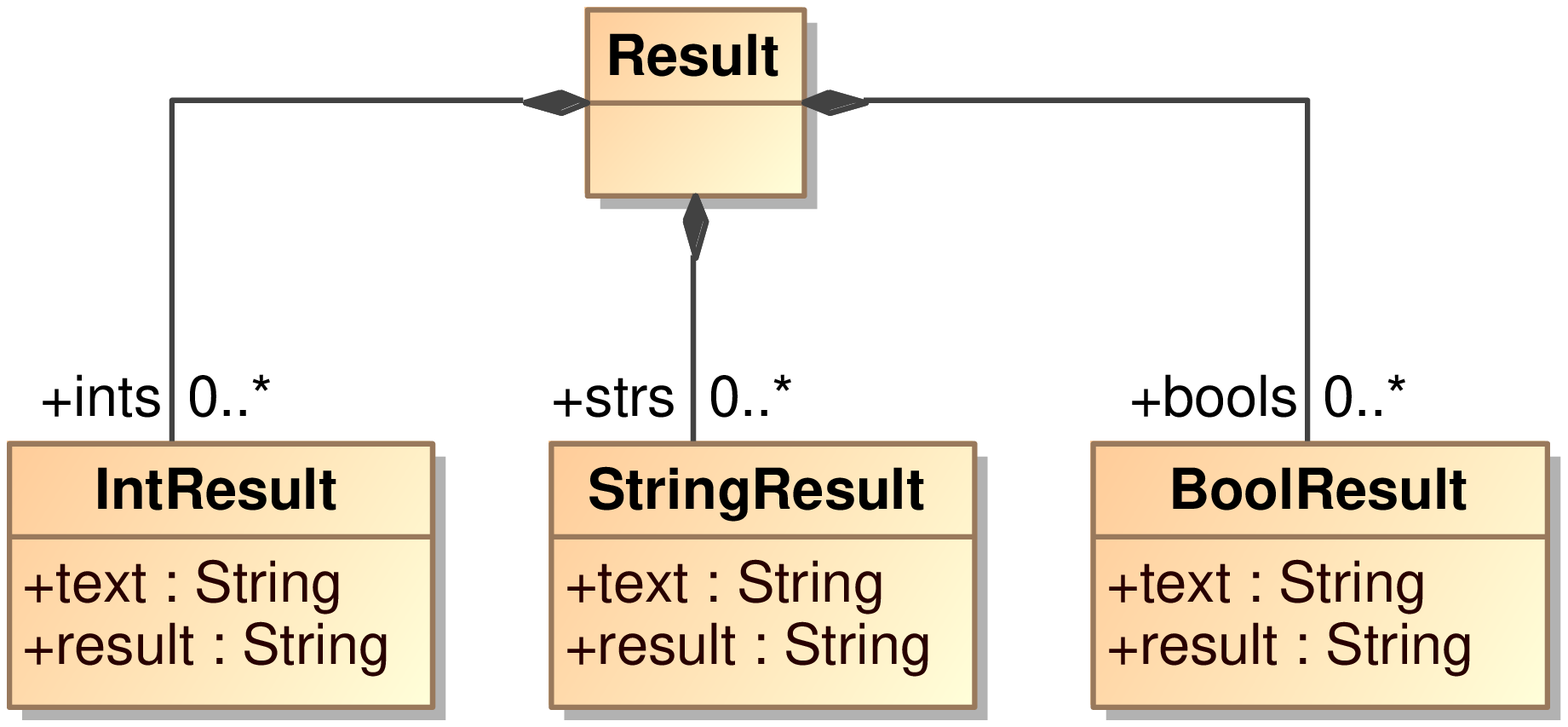}
\end{minipage}%
\\[+2pt]
\begin{minipage}[c]{0.5\linewidth}
\caption{SimpleGraph}
\label{fig:SimpleGraph}
\end{minipage}%
\begin{minipage}[c]{0.5\linewidth}
   \caption{Result}
\label{fig:Result}
\end{minipage}
\end{figure}

\vspace*{-1.5\baselineskip}

\begin{figure}[!h]
\begin{minipage}[t]{0.55\linewidth}
\vspace{0pt} \centering
\includegraphics[width=.9\linewidth]{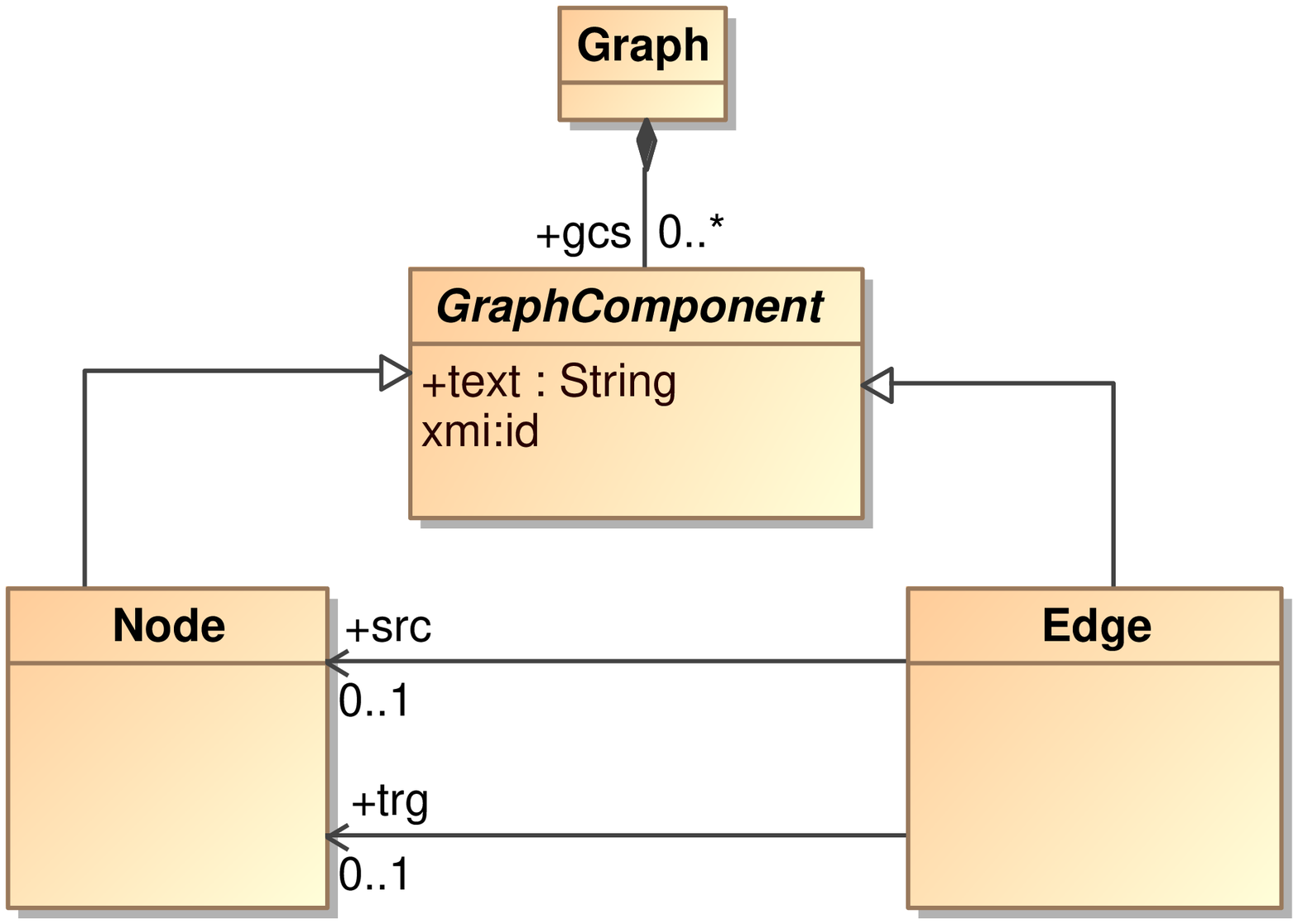}
\end{minipage}%
\begin{minipage}[t]{0.45\linewidth}
\vspace{0pt} \centering
\includegraphics[width=.7\linewidth]{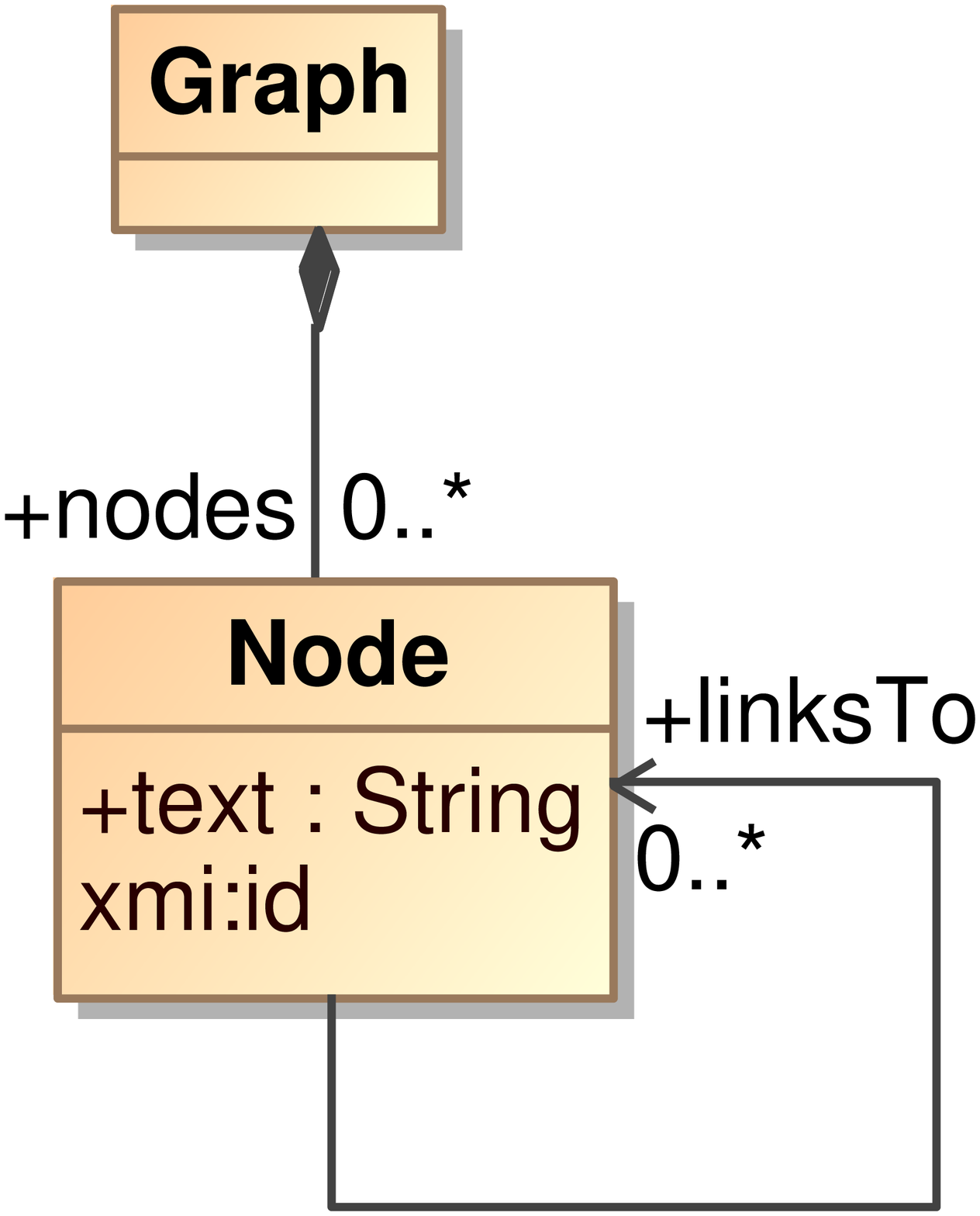}
\end{minipage}%
\\[+2pt]
\begin{minipage}[c]{0.5\linewidth}
\caption{EvolvedGraph}
\label{fig:EvolvedGraph}
\end{minipage}%
\begin{minipage}[c]{0.5\linewidth}
   \caption{MoreEvolvedGraph}
\label{fig:MoreEvolvedGraph}
\end{minipage}
\end{figure}

\section{Transformation for Hello world \label{ap:helloworld}}

\subsection{The constant transformation}

\noindent $\bullet$ \textbf{Configuration:} name : \emph{TTC\_HelloWorld}, source : \emph{SimpleGraph}, sourceKey : \emph{xmi:id}, sourceName : \emph{src}, target: \emph{HelloWorld}, targetKey:\emph{text}, targetName : \emph{trg}.

\vspace*{-1.0\baselineskip}

\begin{figure}[!h]
\begin{minipage}[t]{0.5\linewidth}
\vspace{0pt} \centering
\includegraphics[width=1.0\linewidth]{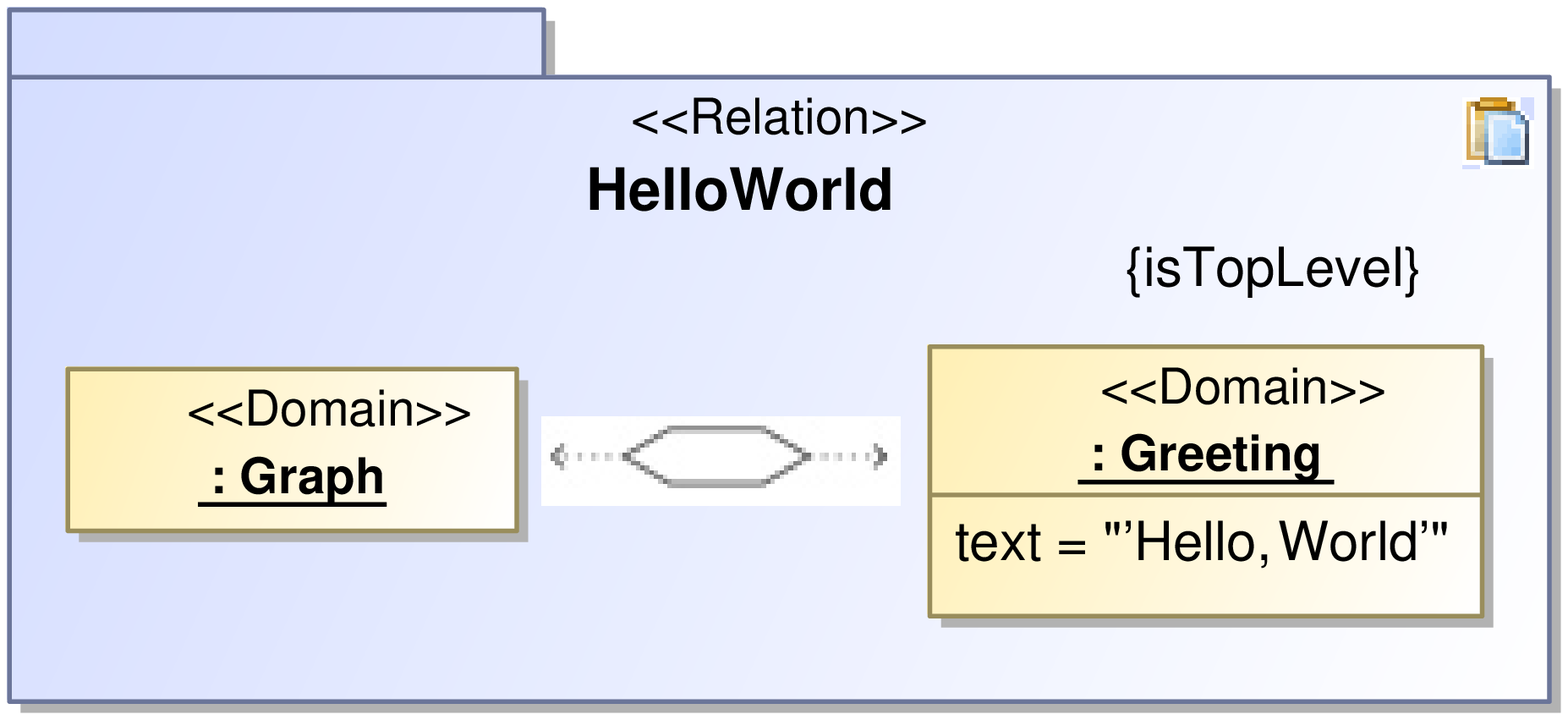}
\end{minipage}%
\begin{minipage}[t]{0.5\linewidth}
\vspace{0pt} \centering
\includegraphics[width=1.0\linewidth]{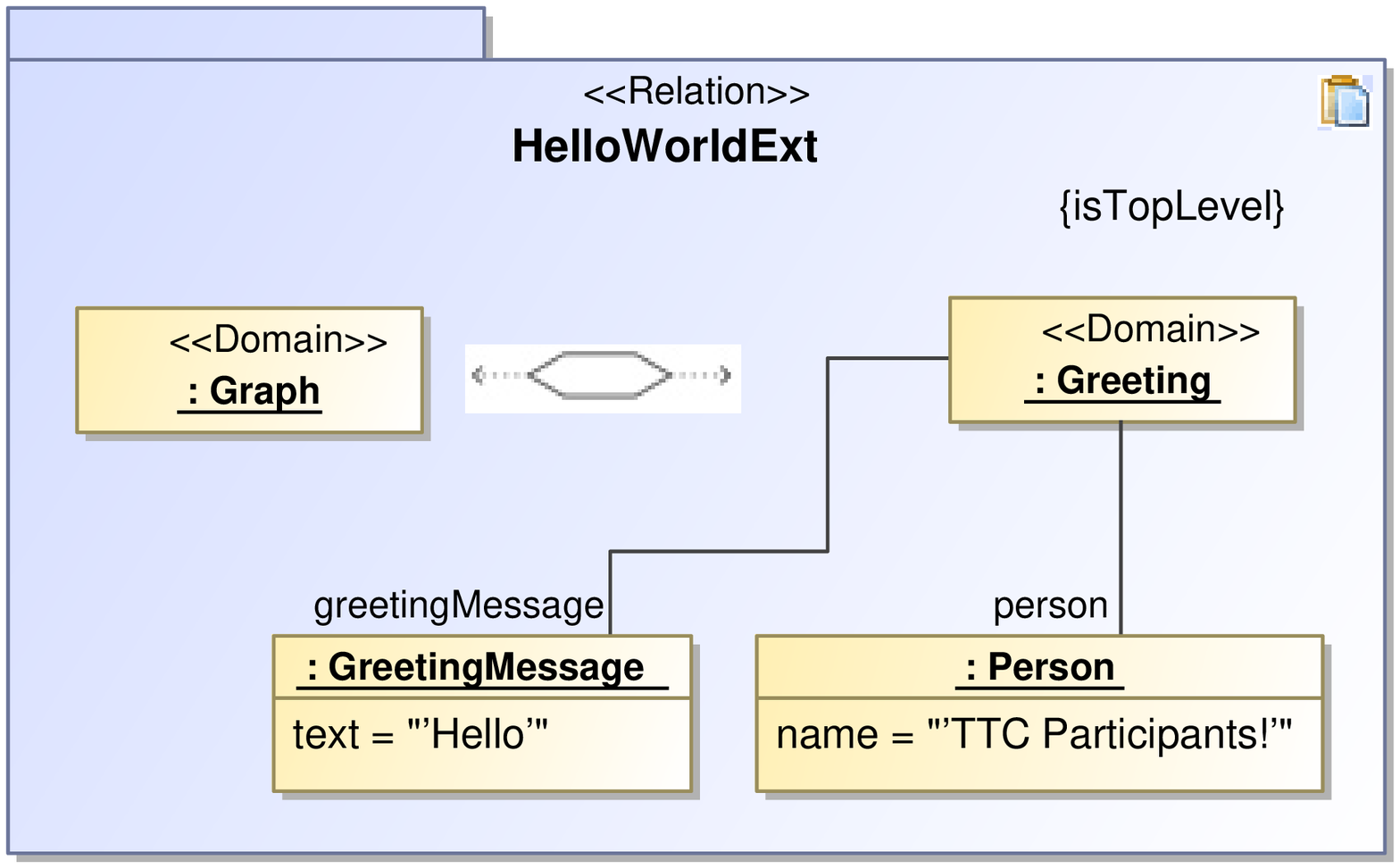}
\end{minipage}%
\\[+2pt]
\begin{minipage}[c]{0.5\linewidth}
\caption{Constant transformation}
\label{fig:HelloWorld}
\end{minipage}%
\begin{minipage}[c]{0.5\linewidth}
   \caption{Constant transformation with extended model}
\label{fig:HelloWorldExt}
\end{minipage}
\end{figure}

\vspace*{-2.5\baselineskip}
\subsection{The constant transformation with  extended model}

\noindent $\bullet$ \textbf{Configuration:} name : \emph{TTC\_HelloWorldExt}, source : \emph{SimpleGraph}, sourceKey : \emph{xmi:id}, sourceName : \emph{src}, target: \emph{HelloWorldExt}, targetKey:\emph{text}, targetName : \emph{trg}.

\subsection{The model-to-text transformation}

\noindent $\bullet$ \textbf{Configuration:} name : \emph{TTC\_HelloWorldText}, output : \emph{html}, source : \emph{HelloWorldExt}, sourceKey : \emph{text}, sourceName : \emph{src}, target: \emph{HtmlMetaModel}, targetKey:\emph{name}, targetName : \emph{html}.

\vspace*{-1.0\baselineskip}
\begin{figure}[!h]
\begin{center}
   \includegraphics[width=0.7\linewidth]{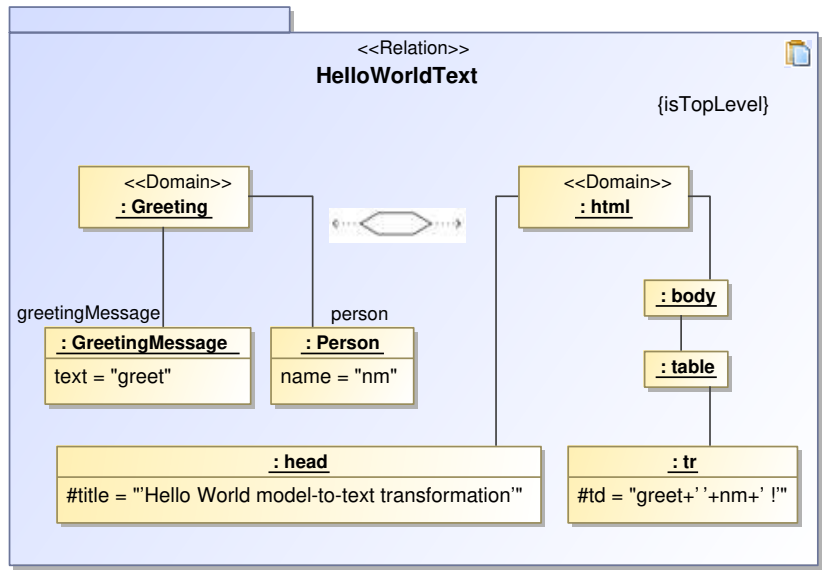}
\end{center}
\vspace*{-1.5\baselineskip}
   \caption{Model-to-text transformation}
\label{fig:HelloWorldText}
\end{figure}

\clearpage
\vspace*{-1.5\baselineskip}
\section{Transformation for Count Matches with Certain Properties \label{ap:countmatches}}
\vspace*{-.5\baselineskip}
\noindent $\bullet$ \textbf{Configuration:} name : \emph{TTC\_CountElement}, source : \emph{SimpleGraph}, sourceKey : \emph{xmi:id}, sourceName : \emph{graph}, target: \emph{Result}, targetKey:\emph{text}, targetName : \emph{result}.

\vspace*{-1.0\baselineskip}
\begin{figure}[!h]
\begin{minipage}[t]{0.6\linewidth}
\vspace{0pt} \centering
\includegraphics[width=.9\linewidth]{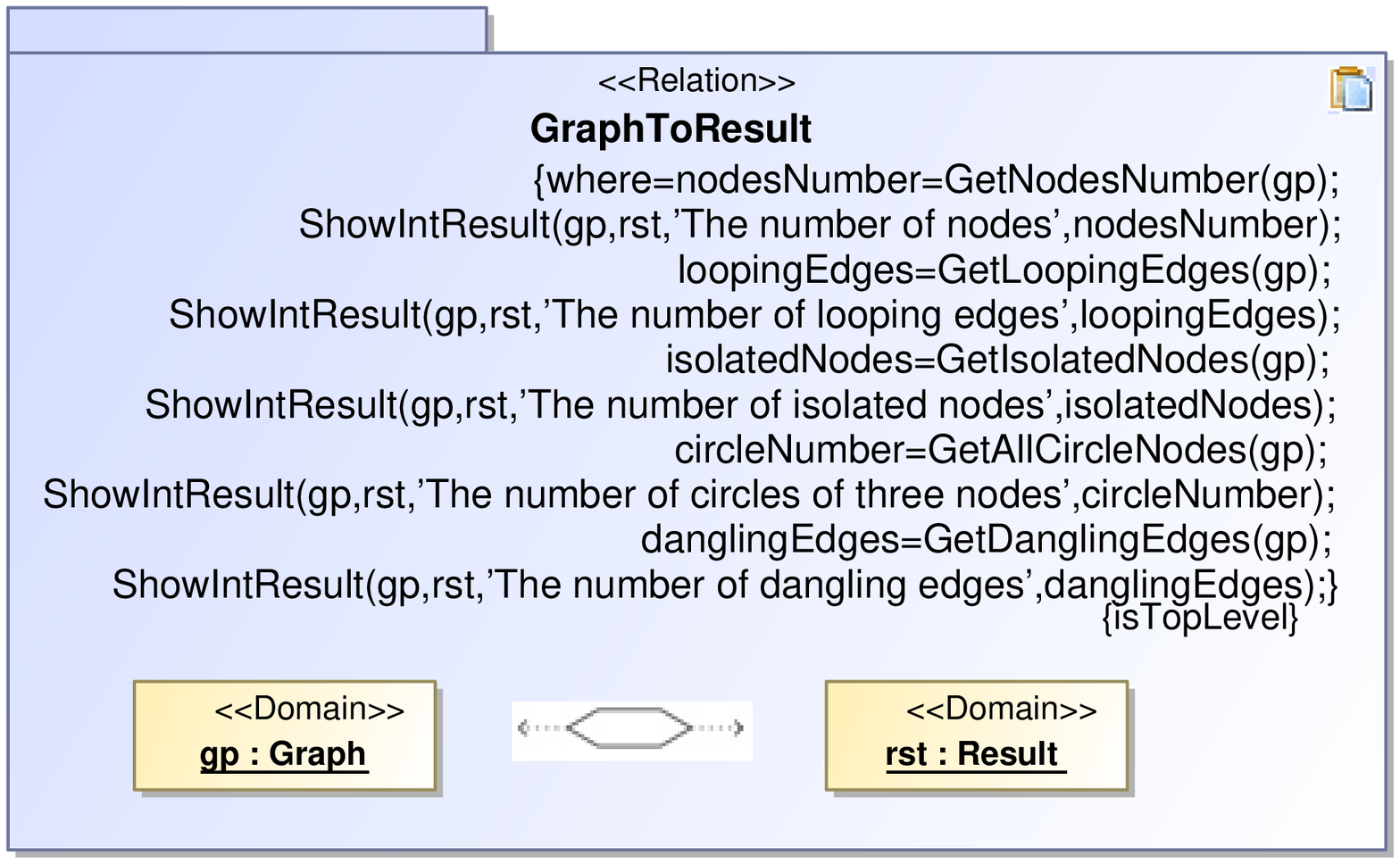}
\end{minipage}%
\begin{minipage}[t]{0.4\linewidth}
\vspace{0pt} \centering
\includegraphics[width=.9\linewidth]{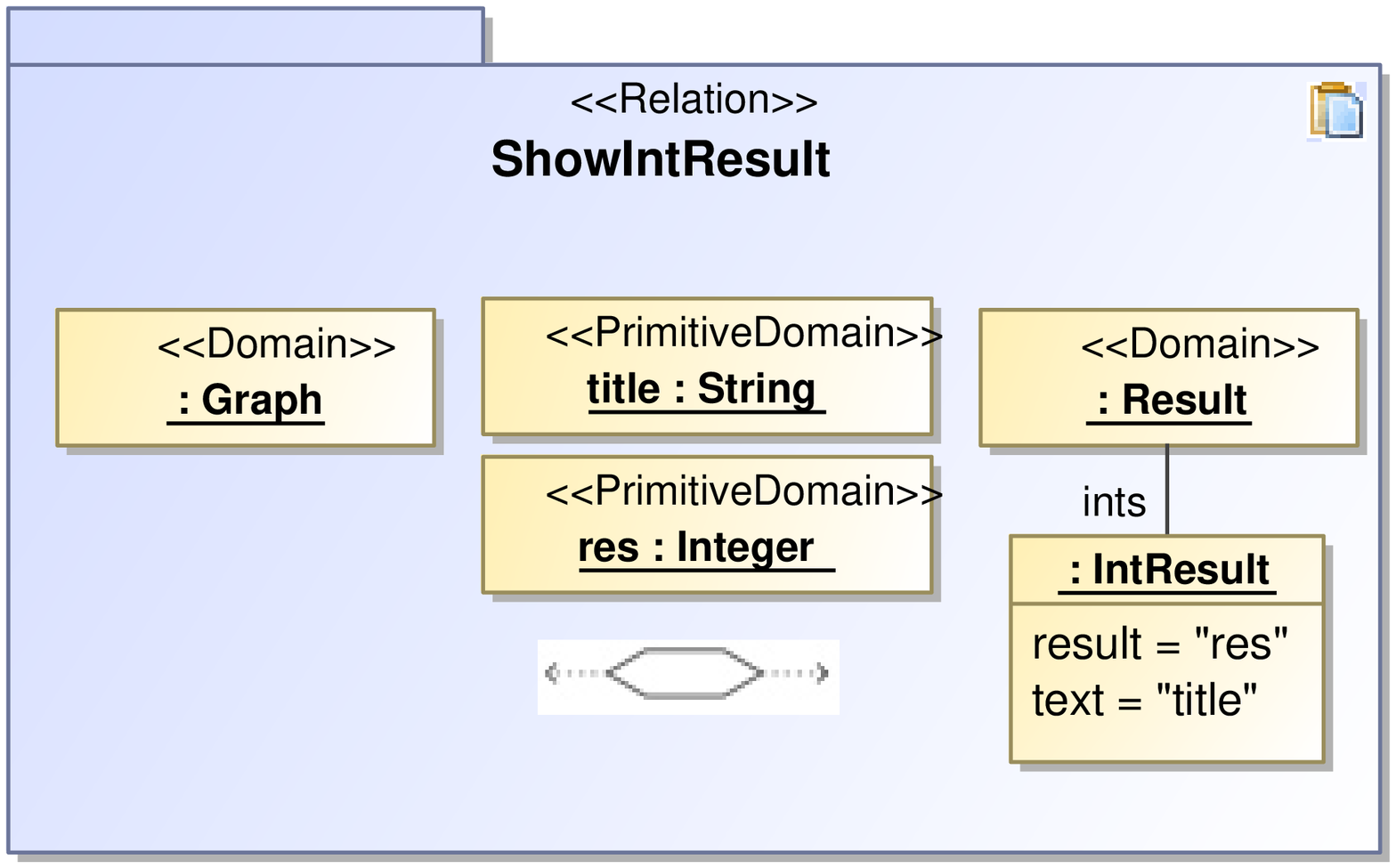}
\end{minipage}%
\\
\vspace*{-1.0\baselineskip}
\begin{minipage}[c]{0.6\linewidth}
\caption{Starting top level relation}
\label{fig:GraphToResult}
\end{minipage}%
\begin{minipage}[c]{0.4\linewidth}
   \caption{Add count result and explanation to result model}
\label{fig:ShowIntResult}
\end{minipage}
\end{figure}

\vspace*{-.5\baselineskip}
\begin{figure}[!h]
\begin{minipage}[t]{0.5\linewidth}
\vspace{0pt} \centering
\includegraphics[width=0.8\linewidth]{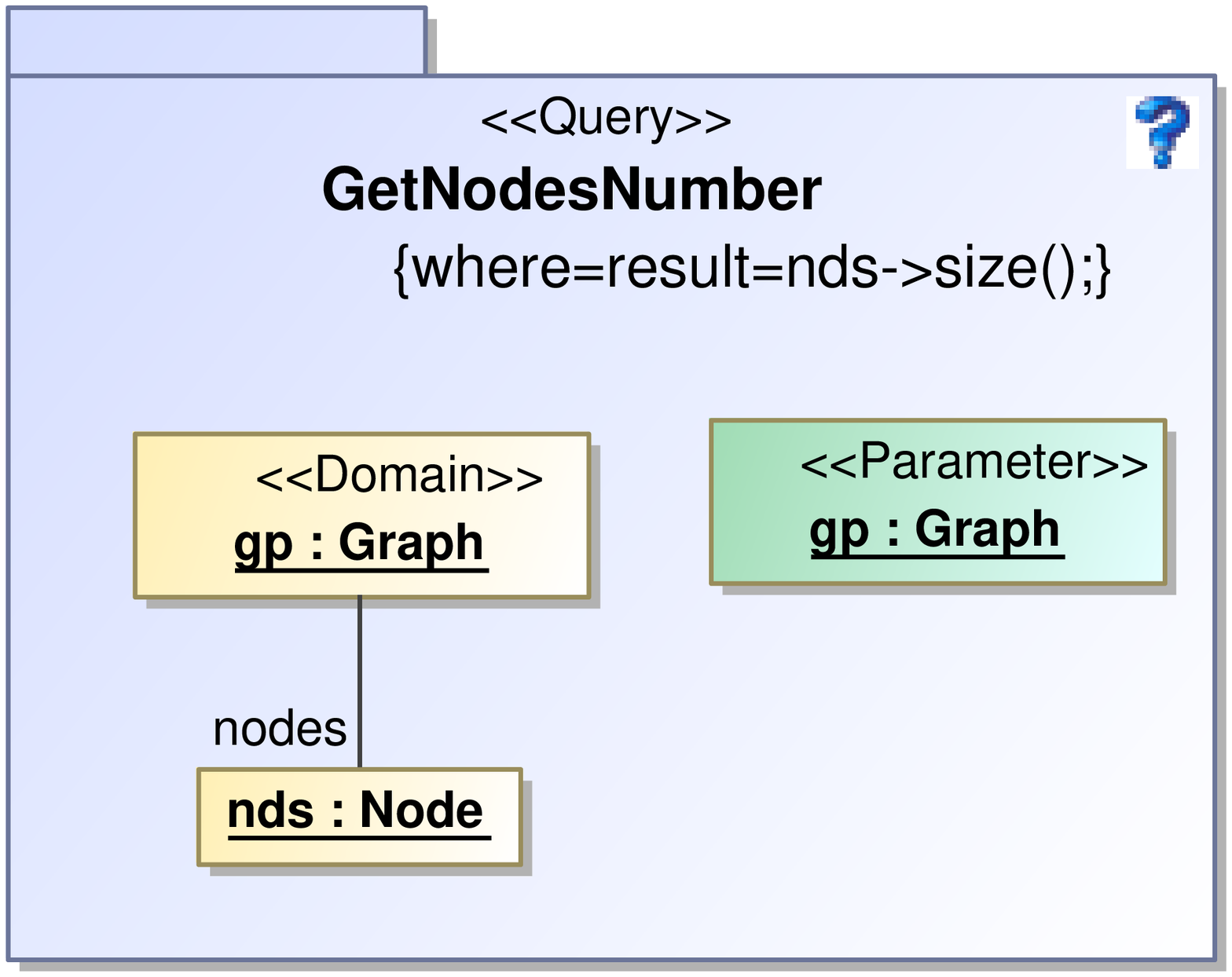}
\end{minipage}%
\begin{minipage}[t]{0.5\linewidth}
\vspace{0pt} \centering
\includegraphics[width=1.0\linewidth]{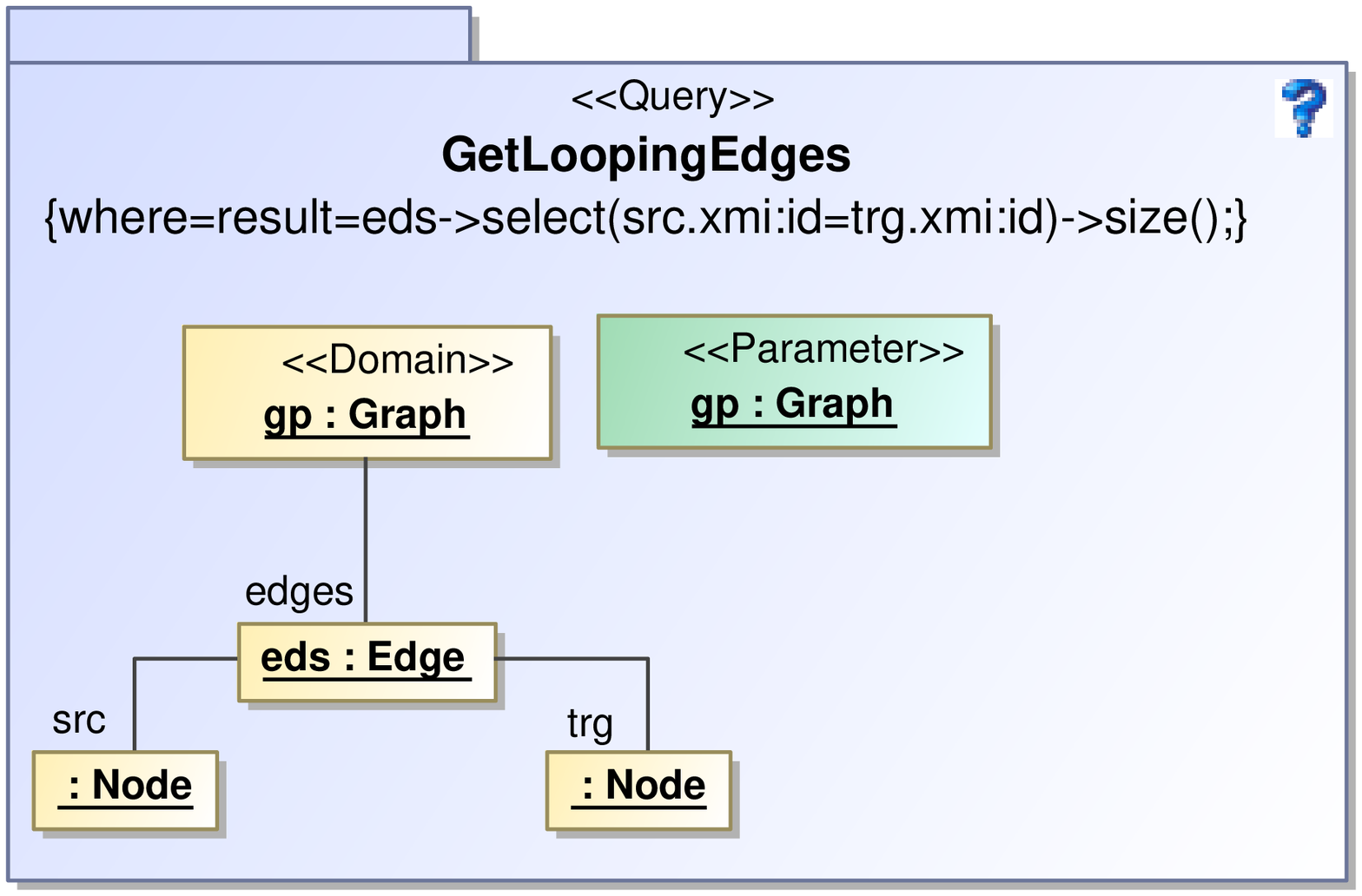}
\end{minipage}%
\\
\vspace*{-2.0\baselineskip}
\begin{minipage}[c]{0.5\linewidth}
\caption{Count the number of nodes}
\label{fig:GetNodesNumber}
\end{minipage}%
\hspace{6pt}
\begin{minipage}[c]{0.5\linewidth}
   \caption{Count the number of looping edges}
\label{fig:GetLoopingEdges}
\end{minipage}%
\\[+25pt]
\begin{minipage}[t]{0.5\linewidth}
\vspace{0pt} \centering
\includegraphics[width=1.0\linewidth]{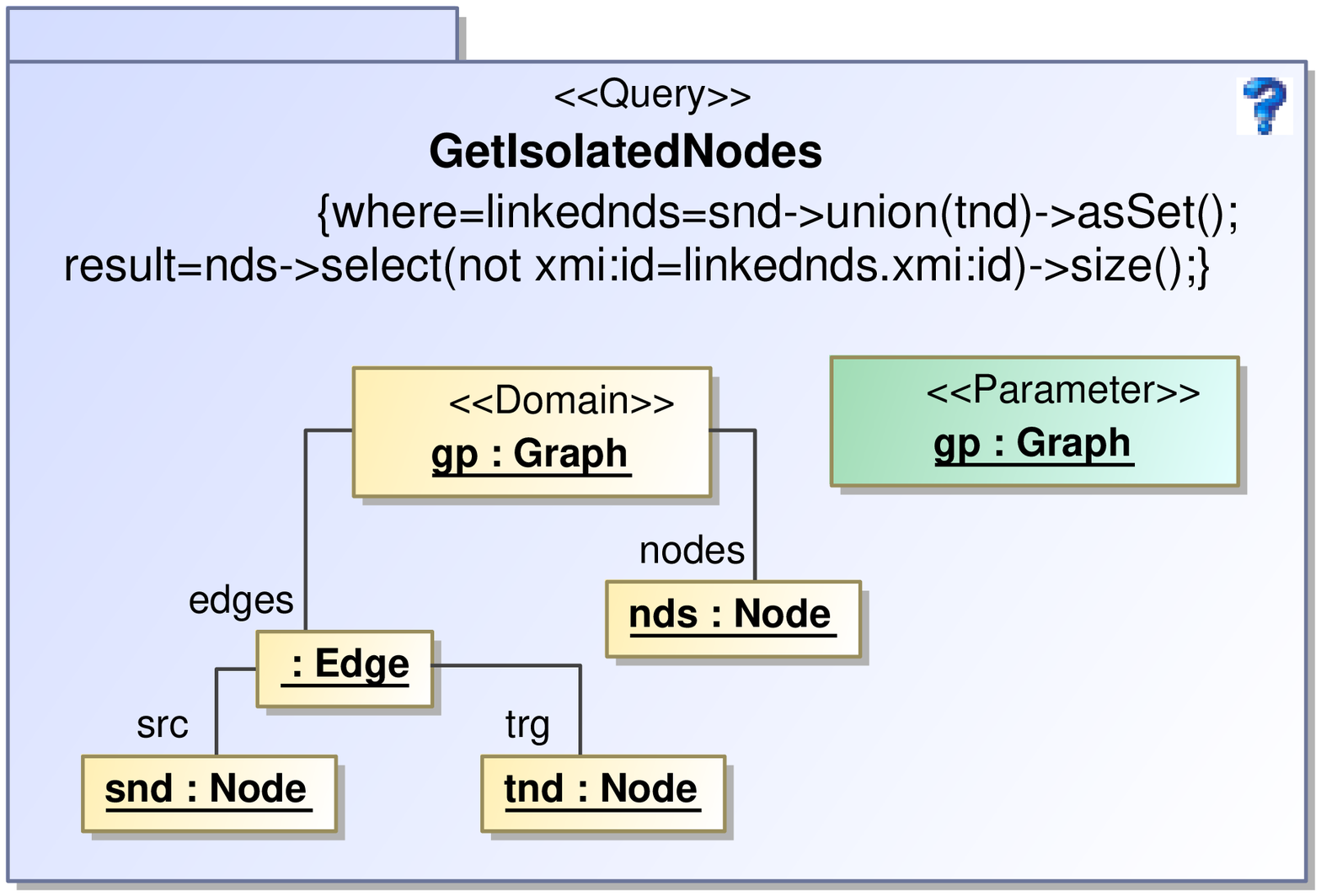}
\end{minipage}%
\begin{minipage}[t]{0.5\linewidth}
\vspace{0pt} \centering
\includegraphics[width=1.0\linewidth]{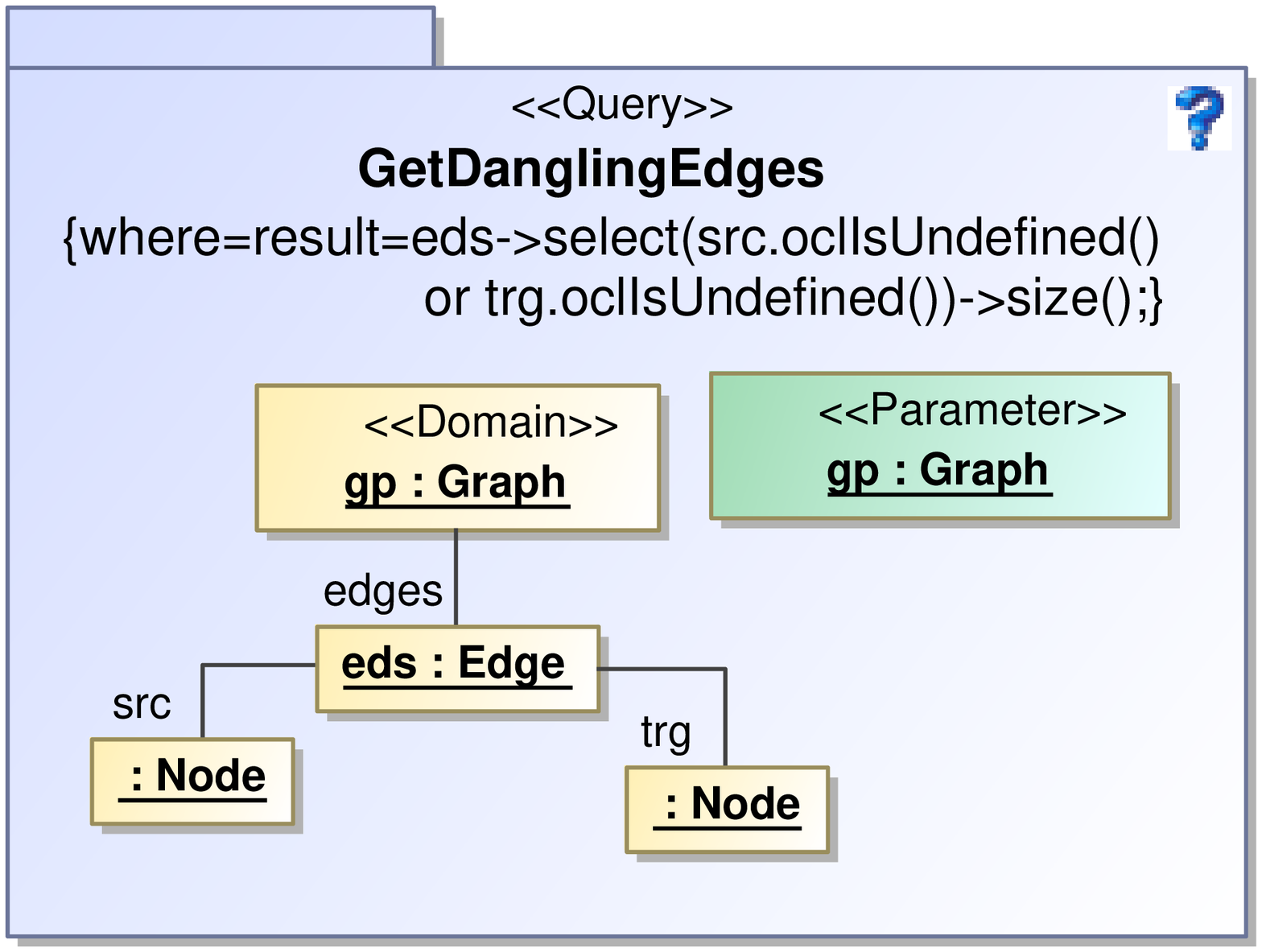}
\end{minipage}%
\\
\vspace*{-2.0\baselineskip}
\begin{minipage}[c]{0.5\linewidth}
\caption{Count the number of isolated nodes}
\label{fig:GetIsolatedNodes}
\end{minipage}%
\hspace{6pt}
\begin{minipage}[c]{0.5\linewidth}
   \caption{Count the number of dangling edges}
\label{fig:GetDanglingEdges}
\end{minipage}%
\end{figure}

\vspace*{-1.5\baselineskip}

\begin{figure}[!h]
\begin{center}
   \includegraphics[width=0.45\linewidth]{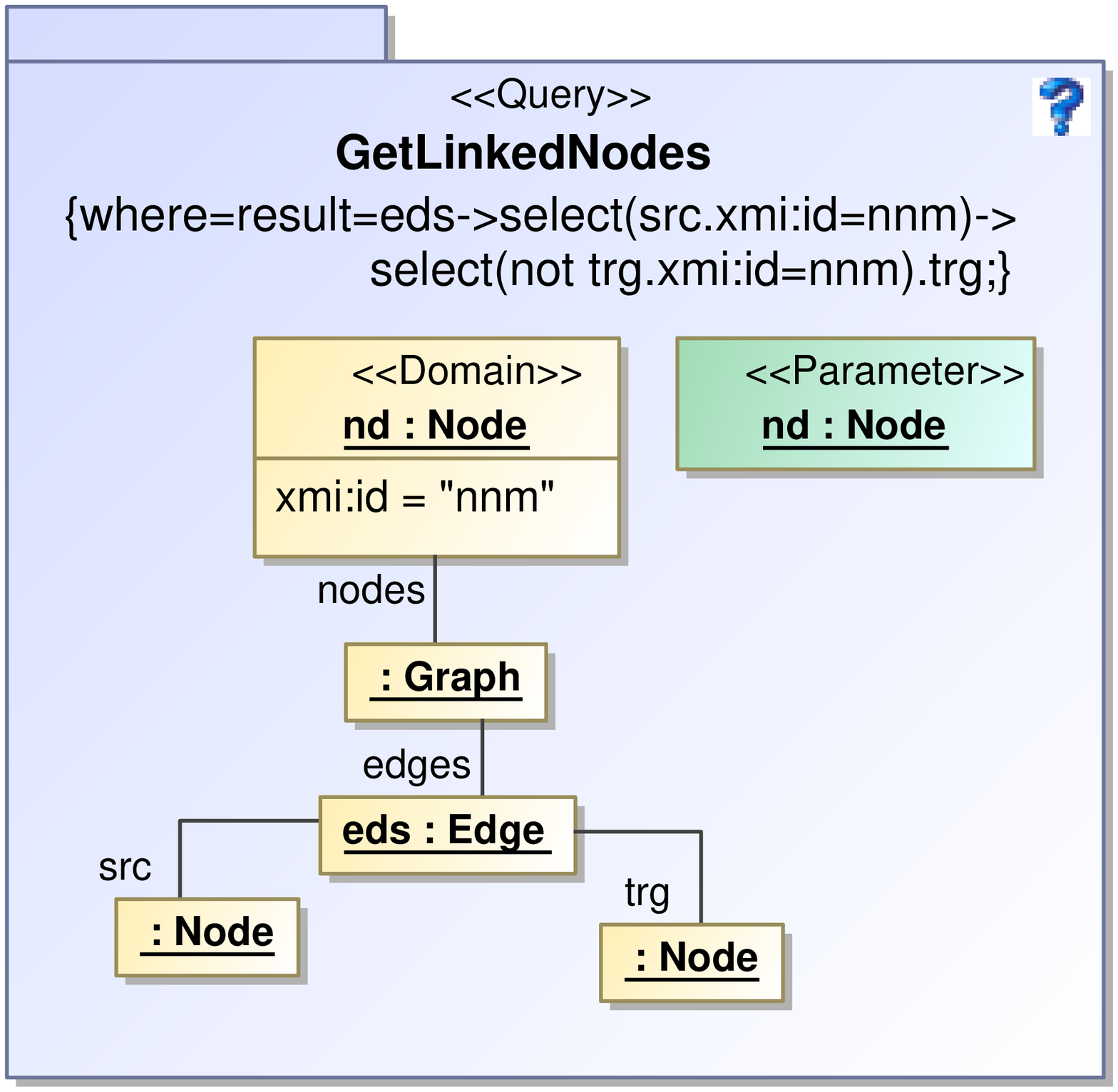}
\end{center}
\vspace*{-1.5\baselineskip}
   \caption{Get target nodes of a node}
\label{fig:GetLinkedNodes}
\end{figure}

\begin{figure}[!h]
\begin{minipage}[c]{1.0\linewidth}
\begin{lstlisting}[language={XSLT},frame=single,numbers=none,basicstyle=\footnotesize,%
morekeywords={}, keywordstyle=\bfseries]
      <xsl:variable name="nds" select="$gp/nodes"/>
      <xsl:variable name="allcnodes" as="item()*">
        <xsl:for-each select="$nds">
            <xsl:sequence select="my:GetCircleNodes(.,.,2)"/>
        </xsl:for-each>
      </xsl:variable>
      <xsl:sequence select="count(distinct-values($allcnodes))"/>
\end{lstlisting}
\caption{Function \textbf{GetAllCircleNodes}(gp : Graph)}
\label{fig:GetAllCircleNodes}
\end{minipage}%
\\[+20pt]
\begin{minipage}[c]{1.0\linewidth}
\begin{lstlisting}[language={XSLT},frame=single,numbers=none,basicstyle=\footnotesize,%
morekeywords={xsl:sort, xsl:sequence}, keywordstyle=\bfseries]
 <xsl:variable name="lnds" select="my:GetLinkedNodes($nd)"/>
 <xsl:for-each select="$lnds">
   <xsl:variable name="cnd" select="."/>
   <xsl:choose>
     <xsl:when test="$counter=0 and $list[1][@xmi:id=$cnd/@xmi:id]">
       <xsl:variable name="rrr" as="item()*">
         <xsl:for-each select="$list">
           <xsl:sort select="@xmi:id" data-type="text" order="ascending"/>
           <val><xsl:value-of select="./@xmi:id"/></val>
         </xsl:for-each>
       </xsl:variable>
       <val><xsl:value-of select="string-join($rrr,'')"/></val>
     </xsl:when>
     <xsl:when test="$counter>0 and not($list[@xmi:id=$cnd/@xmi:id])">
       <xsl:variable name="newlist" select="insert-before($list,count($list)+1,$cnd)"/>
       <xsl:sequence select="my:GetCircleNodes($cnd,$newlist,$counter - 1)"/>
     </xsl:when>
  </xsl:choose>
</xsl:for-each>
\end{lstlisting}
\caption{Function \textbf{GetCircleNodes}(nd: Node, list: Set, counter : Integer): Node}
\label{fig:GetCircleNodes}
\end{minipage}
\end{figure}

\clearpage
\section{Transformation for  Reverse Edges \label{ap:reverseedges}}

\noindent $\bullet$ \textbf{Configuration:} name : \emph{TTC\_ReverseEdges}, source : \emph{SimpleGraph}, sourceKey : \emph{xmi:id}, sourceName : \emph{msrc}, target: \emph{SimpleGraph}, targetKey:\emph{xmi:id}, targetName : \emph{mtrg}.

\vspace*{-1.0\baselineskip}
\begin{figure}[!h]
\begin{center}
   \includegraphics[width=0.45\linewidth]{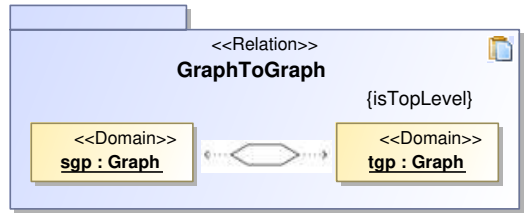}
\end{center}
\vspace*{-1.5\baselineskip}
   \caption{Starting relation---copy a graph}
\label{fig:GraphToGraph_Re}
\end{figure}

\vspace*{-1.5\baselineskip}
\begin{figure}[!h]
\begin{minipage}[t]{0.55\linewidth}
\vspace{0pt} \centering
\includegraphics[width=1.0\linewidth]{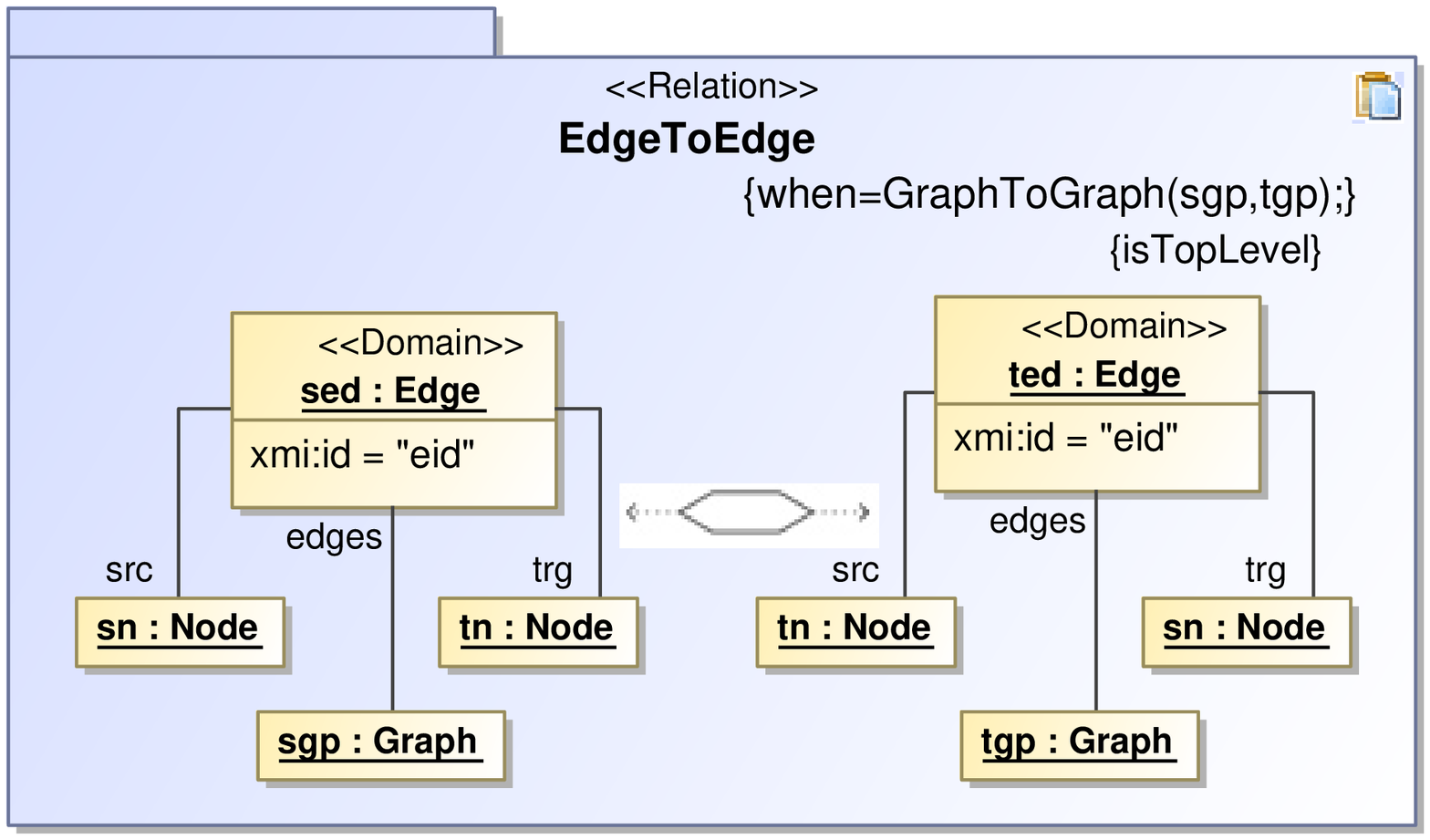}
\end{minipage}%
\begin{minipage}[t]{0.45\linewidth}
\vspace{0pt} \centering
\includegraphics[width=.85\linewidth]{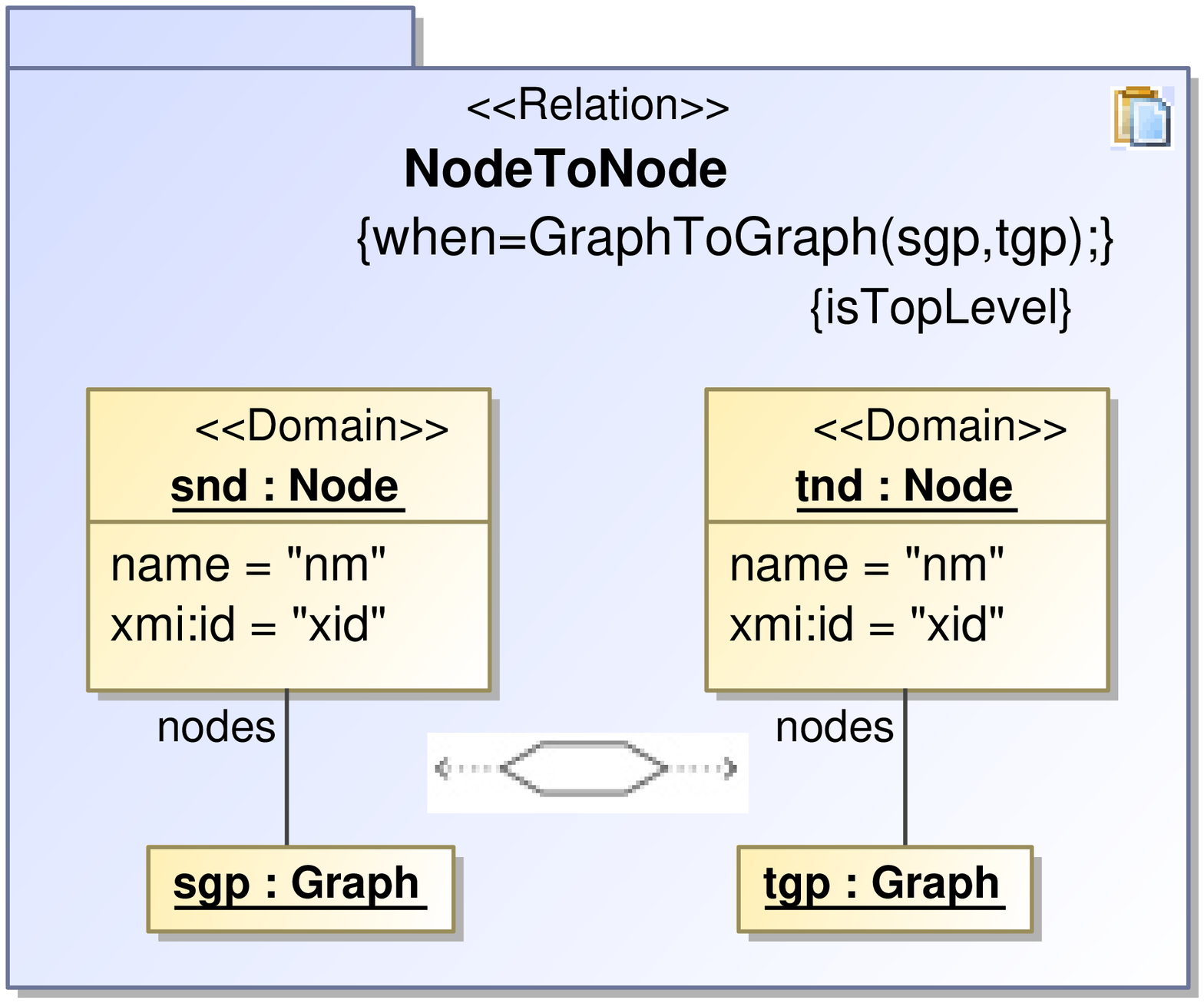}
\end{minipage}%
\\
\vspace*{-1.5\baselineskip}
\begin{minipage}[c]{0.5\linewidth}
\caption{Copy an edge with exchanged source and target nodes}
\label{fig:EdgeToEdge_Re}
\end{minipage}%
\begin{minipage}[c]{0.5\linewidth}
   \caption{Copy a node}
\label{fig:NodeToNode_Re}
\end{minipage}
\end{figure}

\section{Transformation for Simple Migration \label{ap:simplemigration}}

\noindent $\bullet$ \textbf{Configuration:} name : \emph{TTC\_SimpleMigration}, source : \emph{SimpleGraph}, sourceKey : \emph{xmi:id}, sourceName : \emph{msrc}, target: \emph{EvolveGraph}, targetKey:\emph{xmi:id}, targetName : \emph{mtrg}.

\begin{figure}[!h]
\begin{center}
   \includegraphics[width=0.55\linewidth]{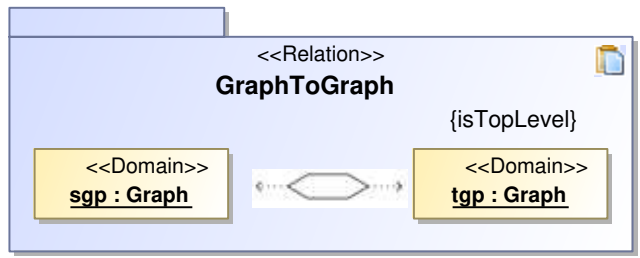}
\end{center}
\vspace*{-1.5\baselineskip}
   \caption{Starting relation---migrate a graph}
\label{fig:GraphToGraph_Sm}
\end{figure}

\clearpage

\begin{figure}[!h]
\begin{minipage}[t]{0.6\linewidth}
\vspace{0pt} \centering
\includegraphics[width=1.0\linewidth]{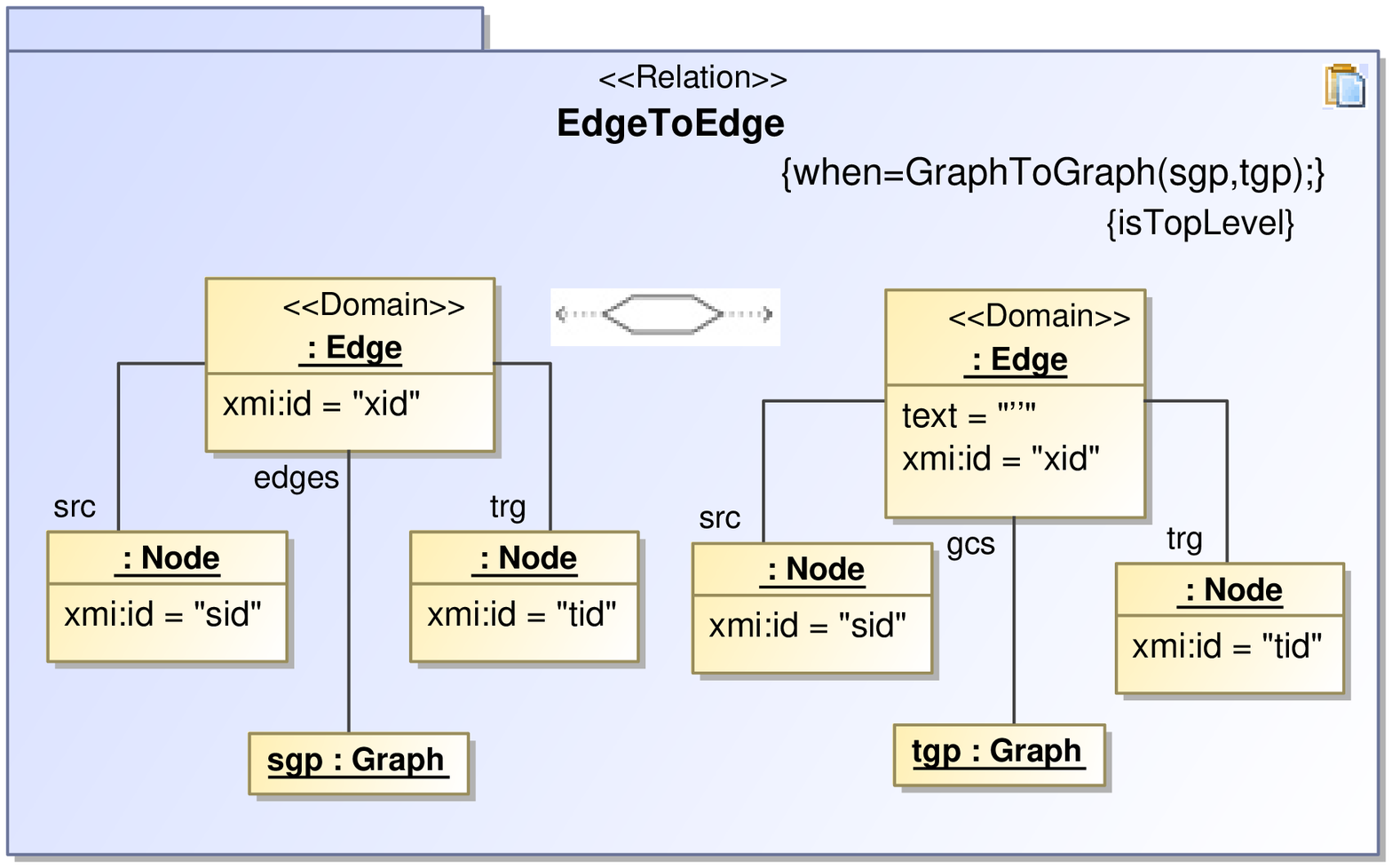}
\end{minipage}%
\begin{minipage}[t]{0.4\linewidth}
\vspace{0pt} \centering
\includegraphics[width=1.0\linewidth]{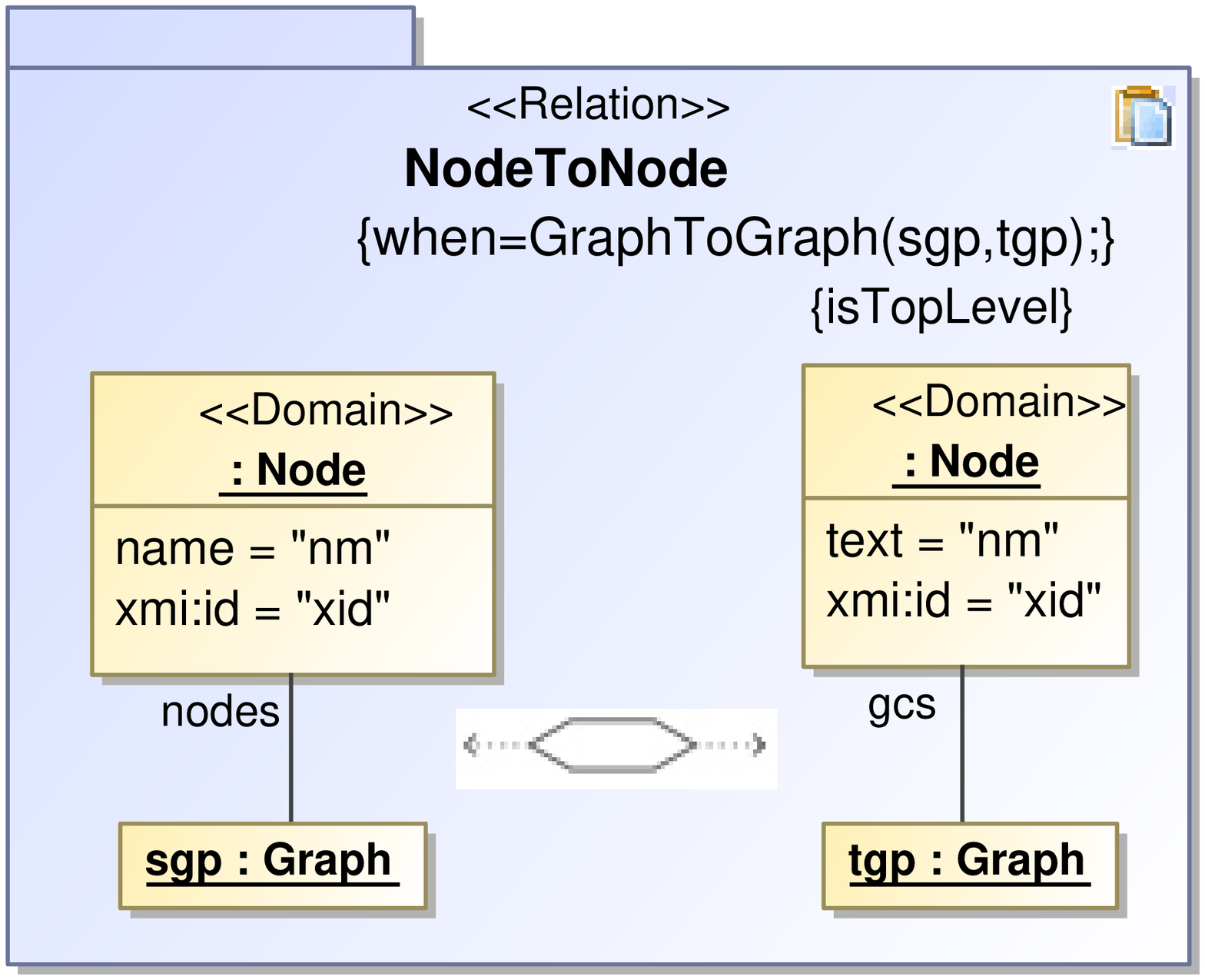}
\end{minipage}%
\\
\vspace*{-1.5\baselineskip}
\begin{minipage}[c]{0.5\linewidth}
\caption{Migrate an edge}
\label{fig:EdgeToEdge_Sm}
\end{minipage}%
\begin{minipage}[c]{0.5\linewidth}
   \caption{Migrate a node}
\label{fig:NodeToNode_Sm}
\end{minipage}
\end{figure}

\section{Transformation for topology-changing migration \label{ap:topology}}

\noindent $\bullet$ \textbf{Configuration:} name : \emph{TTC\_TopologyMigration}, source : \emph{SimpleGraph}, sourceKey : \emph{xmi:id}, sourceName : \emph{msrc}, target: \emph{MoreEvolveGraph}, targetKey:\emph{xmi:id}, targetName : \emph{mtrg}.

\vspace*{-1.0\baselineskip}
\begin{figure}[!h]
\begin{minipage}[t]{0.45\linewidth}
\vspace{0pt} \centering
\includegraphics[width=.9\linewidth]{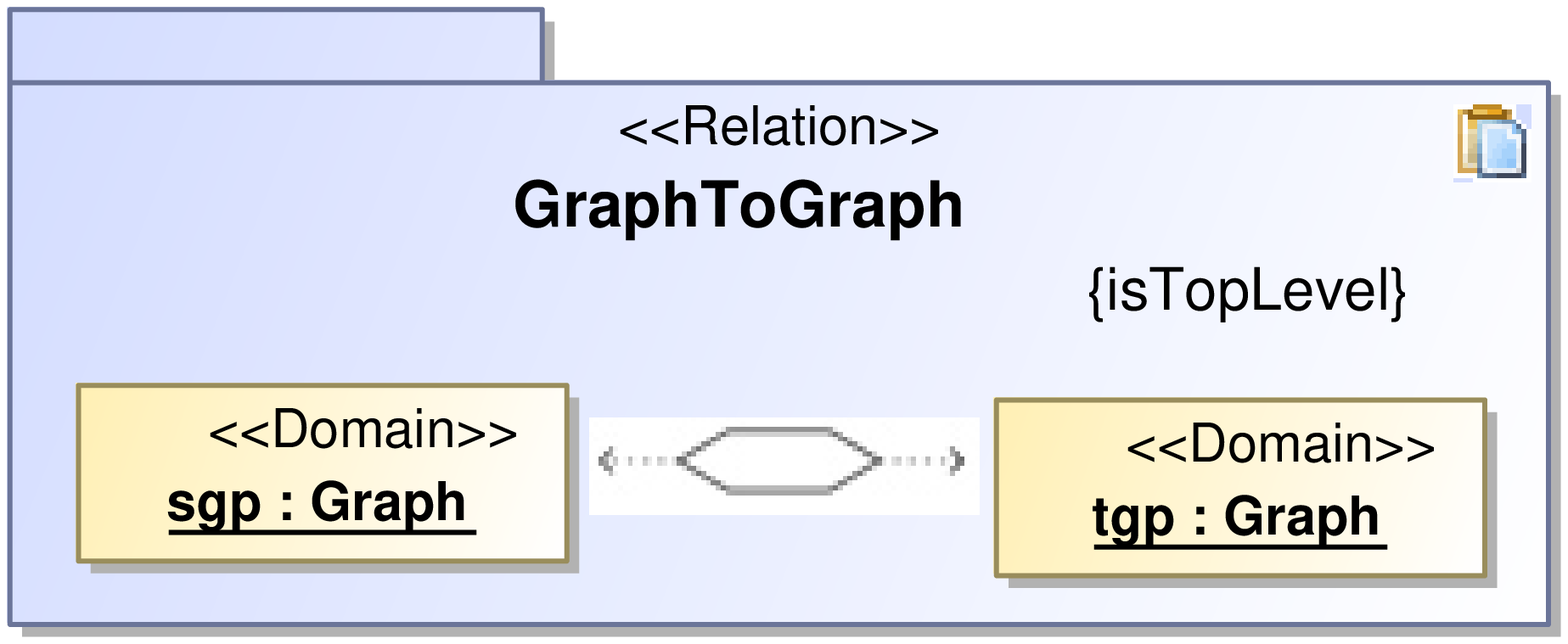}
\end{minipage}%
\begin{minipage}[t]{0.55\linewidth}
\vspace{0pt} \centering
\includegraphics[width=0.8\linewidth]{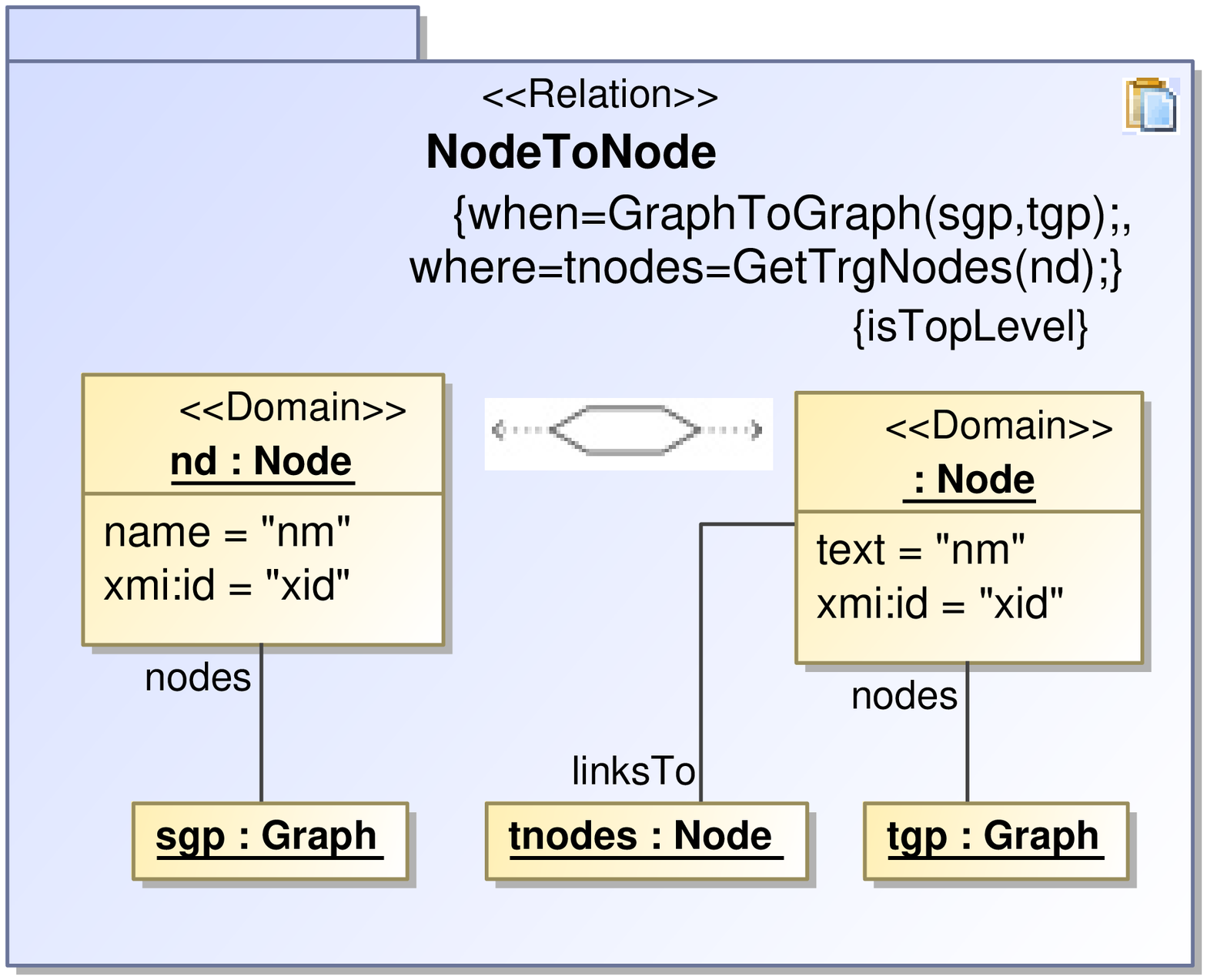}
\end{minipage}%
\\
\vspace*{-1.5\baselineskip}
\begin{minipage}[c]{0.5\linewidth}
\caption{Starting relation---migrate a graph}
\label{fig:GraphToGraph_Tm}
\end{minipage}%
\begin{minipage}[c]{0.5\linewidth}
   \caption{Migrate a node}
\label{fig:NodeToNode_Tm}
\end{minipage}
\end{figure}

\vspace*{-.5\baselineskip}
\begin{figure}[!h]
\begin{center}
   \includegraphics[width=0.40\linewidth]{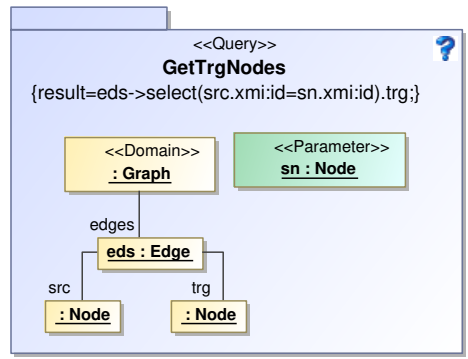}
\end{center}
\vspace*{-1.5\baselineskip}
   \caption{Get target nodes of a node}
\label{fig:GetTrgNodes}
\end{figure}

\section{Transformation for  Delete Node \label{ap:deletenode}}

\noindent $\bullet$ \textbf{Configuration:} name : \emph{TTC\_DeleteNode}, isInPlace : \emph{true}, source : \emph{SimpleGraph}, sourceKey : \emph{xmi:id}, sourceName : \emph{msrc}, target: \emph{SimpleGraph}, targetKey:\emph{xmi:id}, targetName : \emph{mtrg}.

\vspace*{-1.0\baselineskip}
\begin{figure}[!h]
\begin{minipage}[t]{0.5\linewidth}
\vspace{0pt} \centering
\includegraphics[width=1.0\linewidth]{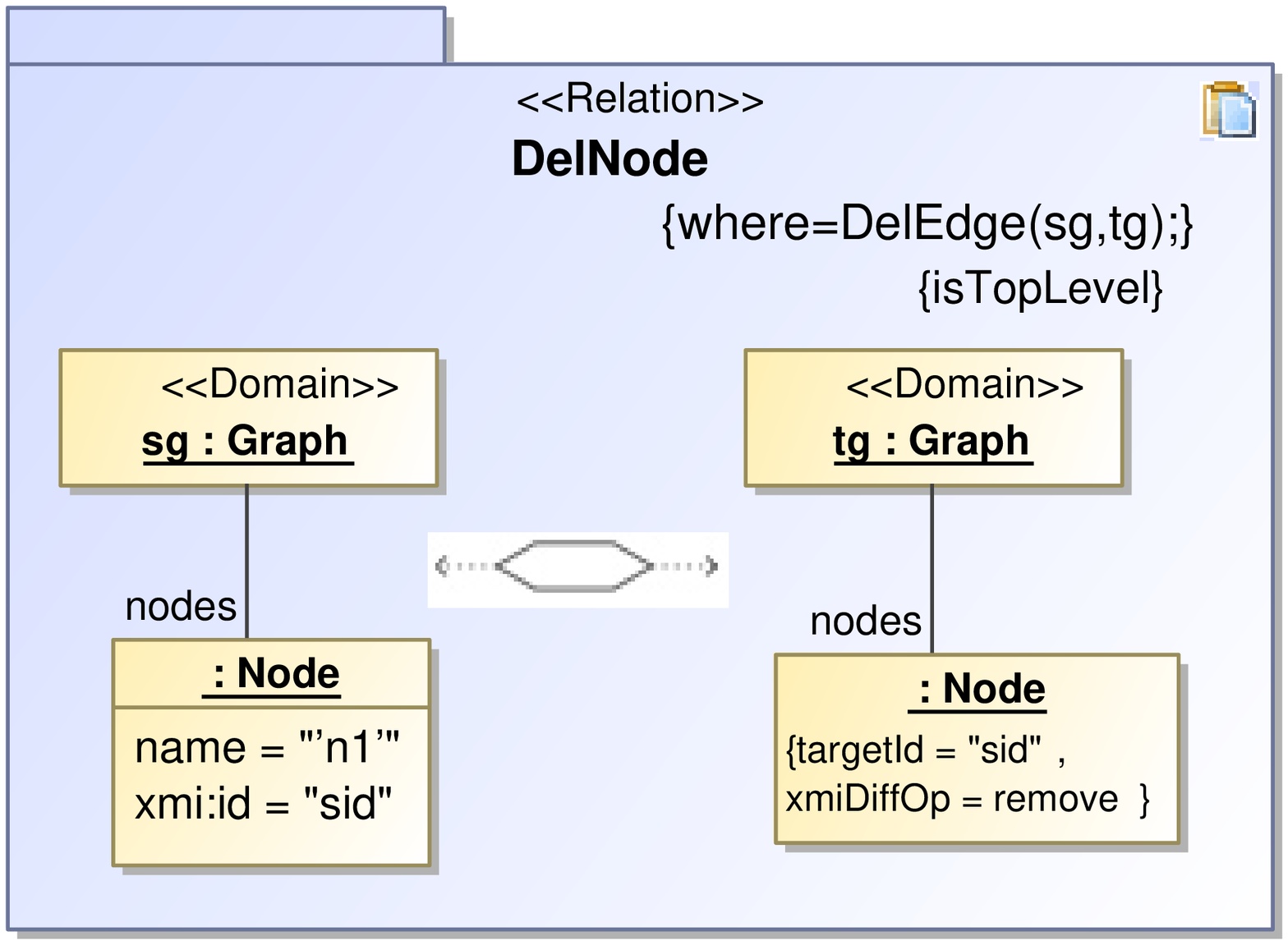}
\end{minipage}%
\begin{minipage}[t]{0.5\linewidth}
\vspace{0pt} \centering
\includegraphics[width=0.8\linewidth]{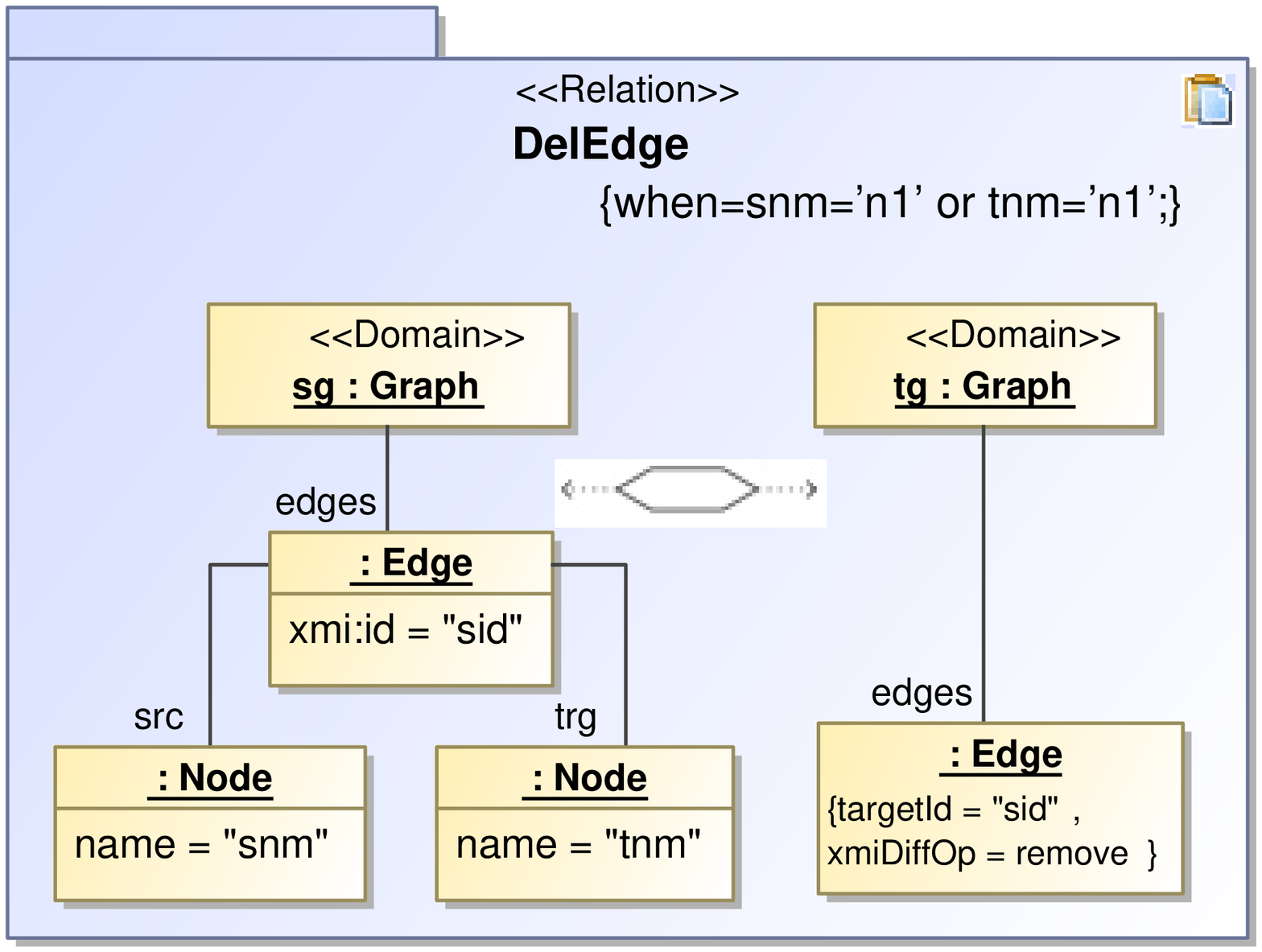}
\end{minipage}%
\\
\vspace*{-1.5\baselineskip}
\begin{minipage}[c]{0.5\linewidth}
\caption{Delete a node with name "n1"}
\label{fig:DelNode}
\end{minipage}%
\begin{minipage}[c]{0.5\linewidth}
   \caption{Delete the incident edges}
\label{fig:DelEdge}
\end{minipage}
\end{figure}

\section{Transformation for  Insert Transitive Edges \label{ap:insertedges}}

\noindent $\bullet$ \textbf{Configuration:} name : \emph{TTC\_InsertTransitiveEdges}, source : \emph{SimpleGraph}, sourceKey : \emph{xmi:id}, sourceName : \emph{msrc}, target: \emph{SimpleGraph}, targetKey:\emph{xmi:id}, targetName : \emph{mtrg}.

\vspace*{-1.0\baselineskip}
\begin{figure}[!h]
\begin{minipage}[t]{0.45\linewidth}
\vspace{0pt} \centering
\includegraphics[width=1.0\linewidth]{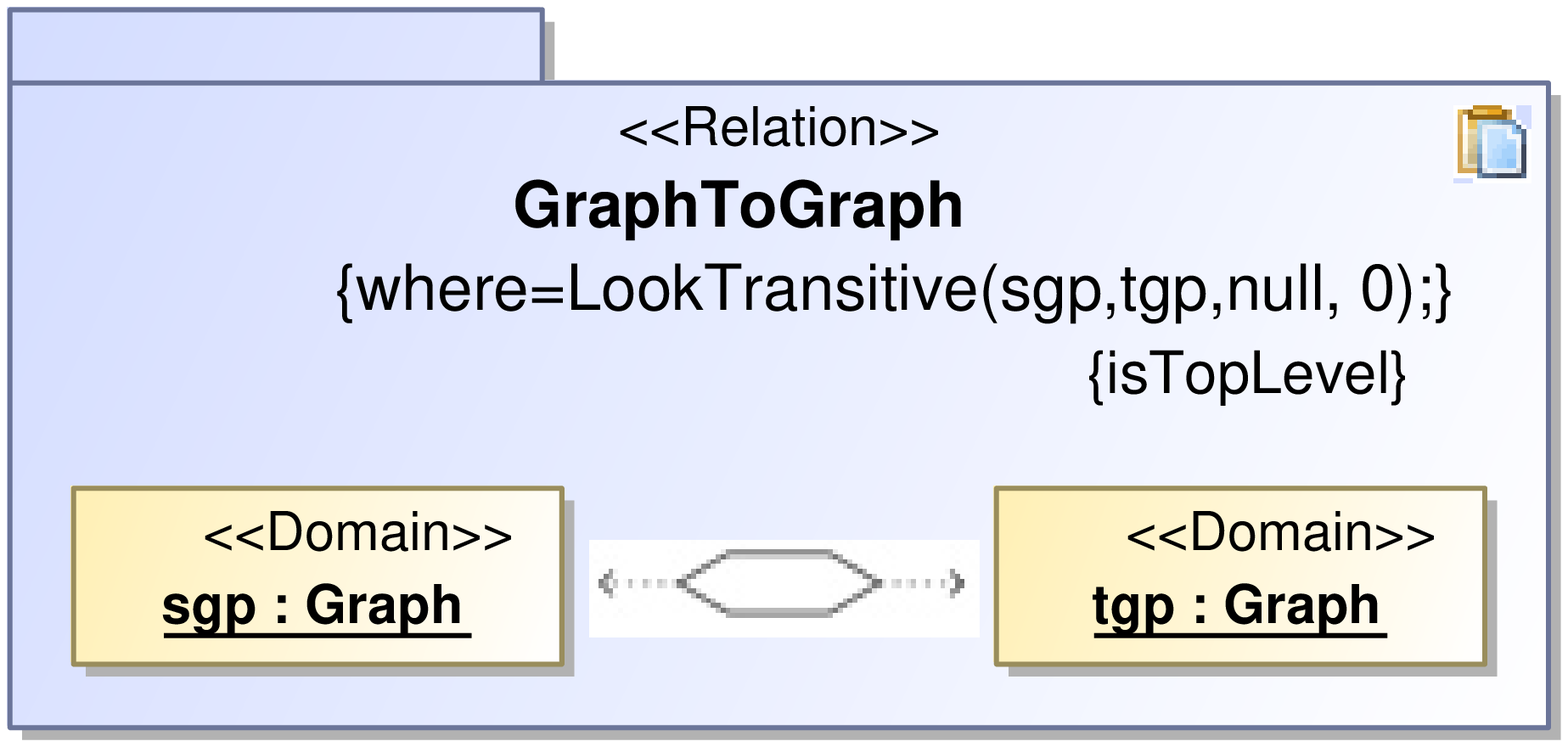}
\end{minipage}%
\begin{minipage}[t]{0.5\linewidth}
\vspace{0pt} \centering
\includegraphics[width=0.8\linewidth]{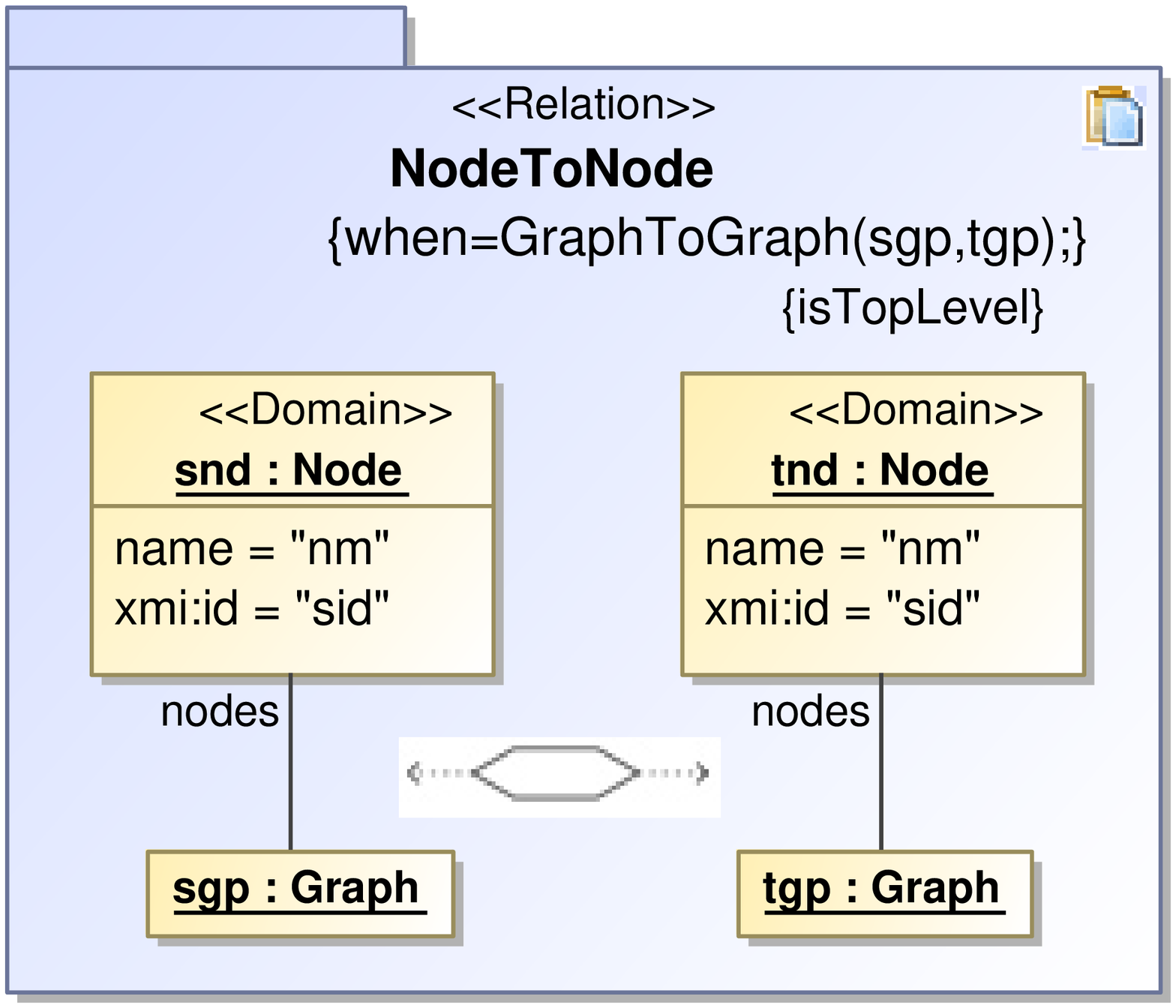}
\end{minipage}%
\\
\vspace*{-1.5\baselineskip}
\begin{minipage}[c]{0.5\linewidth}
\caption{Starting relation---copy a graph}
\label{fig:GraphToGraph_It}
\end{minipage}%
\begin{minipage}[c]{0.5\linewidth}
   \caption{Copy a node}
\label{fig:NodeToNode_It}
\end{minipage}
\end{figure}

\vspace*{-1.5\baselineskip}
\begin{figure}[!h]
\begin{center}
   \includegraphics[width=0.85\linewidth]{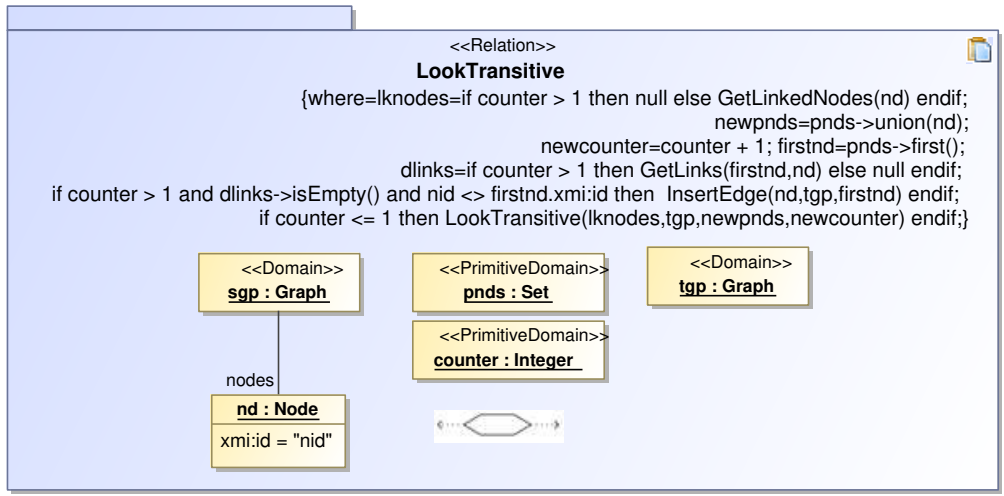}
\end{center}
\vspace*{-1.5\baselineskip}
   \caption{Select two nodes for inserting an additional edge}
\label{fig:LookTransitive}
\end{figure}

\vspace*{-1.5\baselineskip}
\begin{figure}[!h]
\begin{minipage}[t]{0.5\linewidth}
\vspace{0pt} \centering
\includegraphics[width=1.0\linewidth]{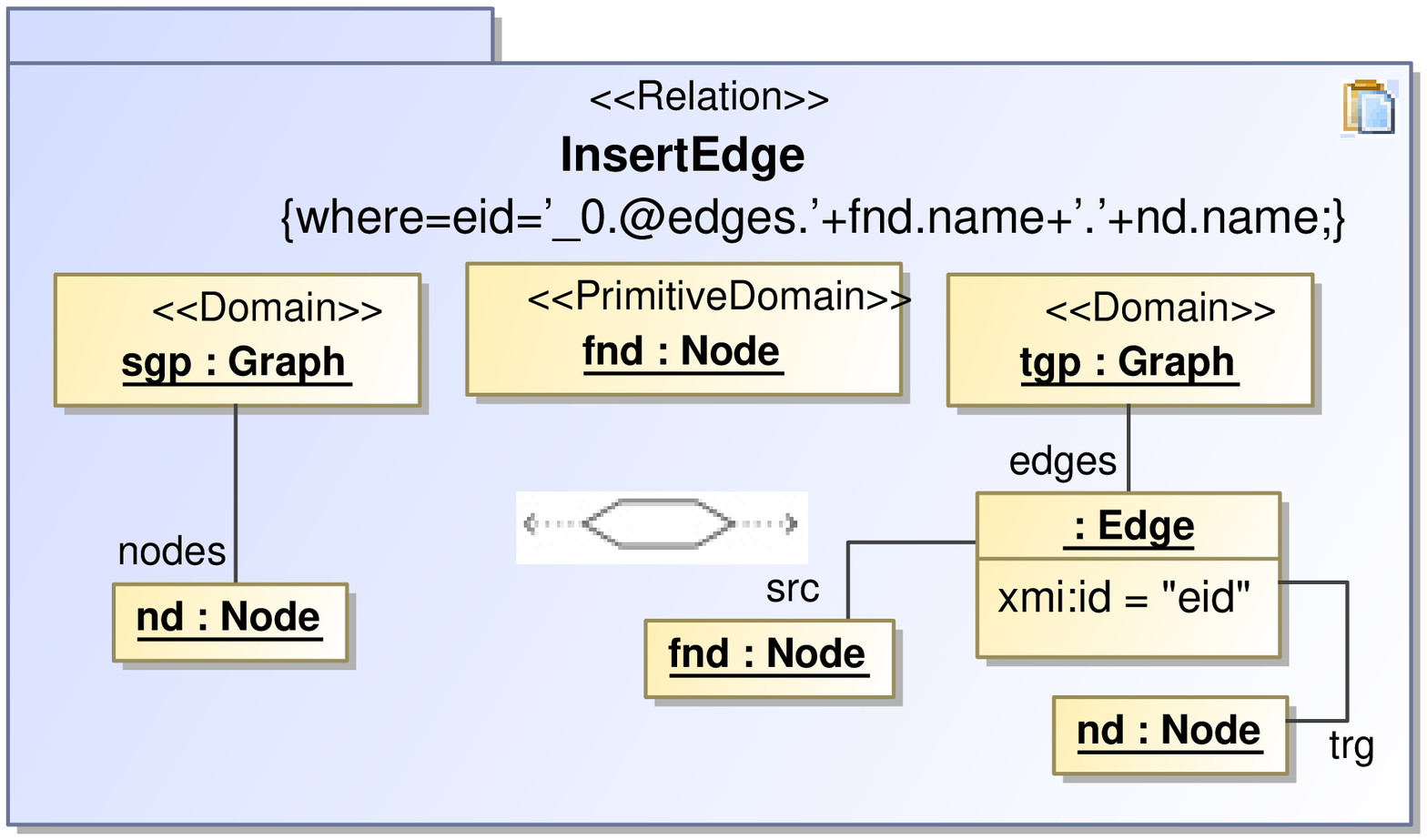}
\end{minipage}%
\begin{minipage}[t]{0.55\linewidth}
\vspace{0pt} \centering
\includegraphics[width=1.0\linewidth]{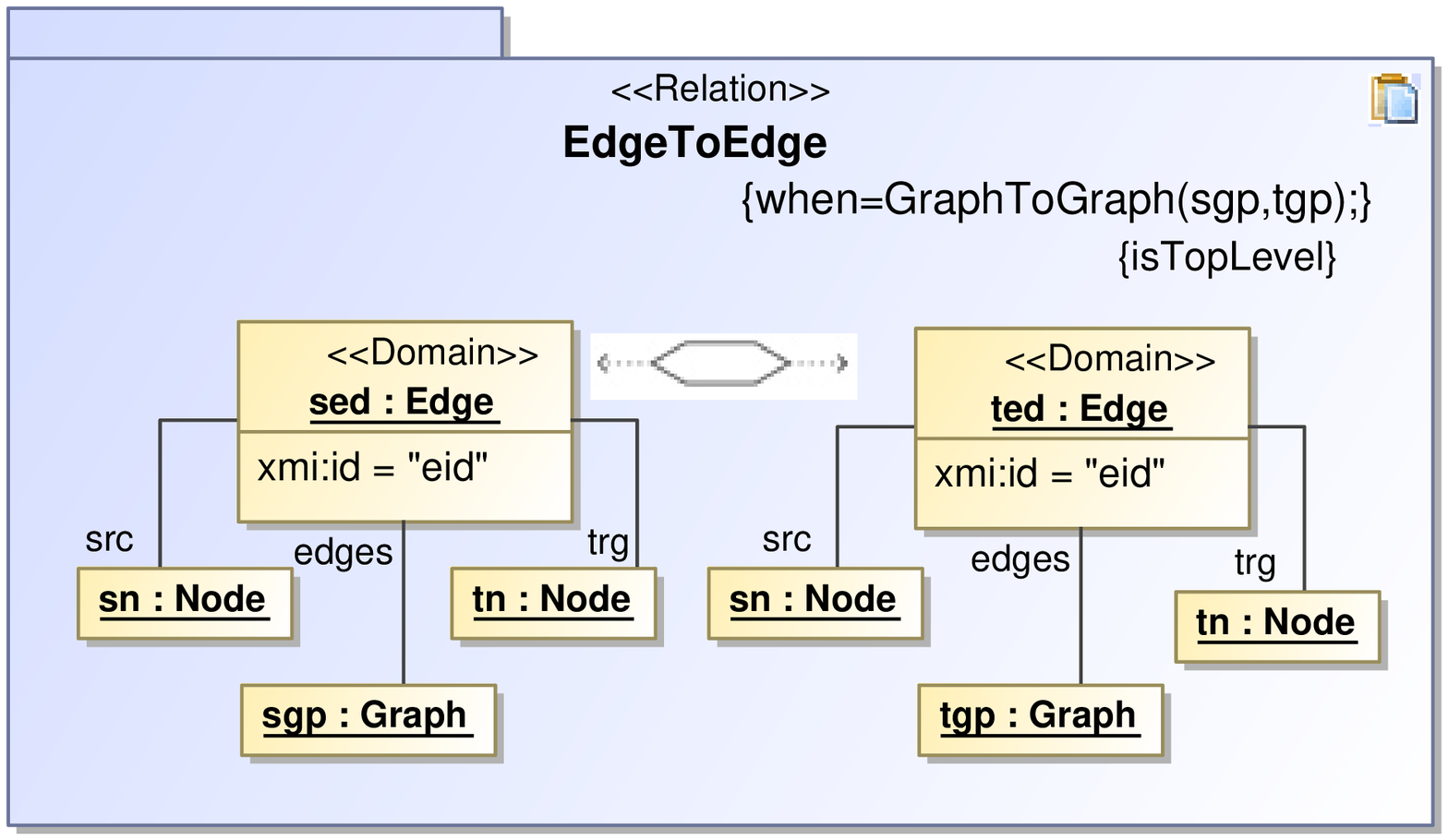}
\end{minipage}%
\\
\vspace*{-1.5\baselineskip}
\begin{minipage}[c]{0.5\linewidth}
\caption{Insert a new edge}
\label{fig:InsertEdge}
\end{minipage}%
\begin{minipage}[c]{0.5\linewidth}
   \caption{Copy an edge}
\label{fig:EdgeToEdge_It}
\end{minipage}
\end{figure}

\vspace*{-1.0\baselineskip}
\begin{figure}[!h]
\begin{minipage}[t]{0.5\linewidth}
\vspace{0pt} \centering
\includegraphics[width=.9\linewidth]{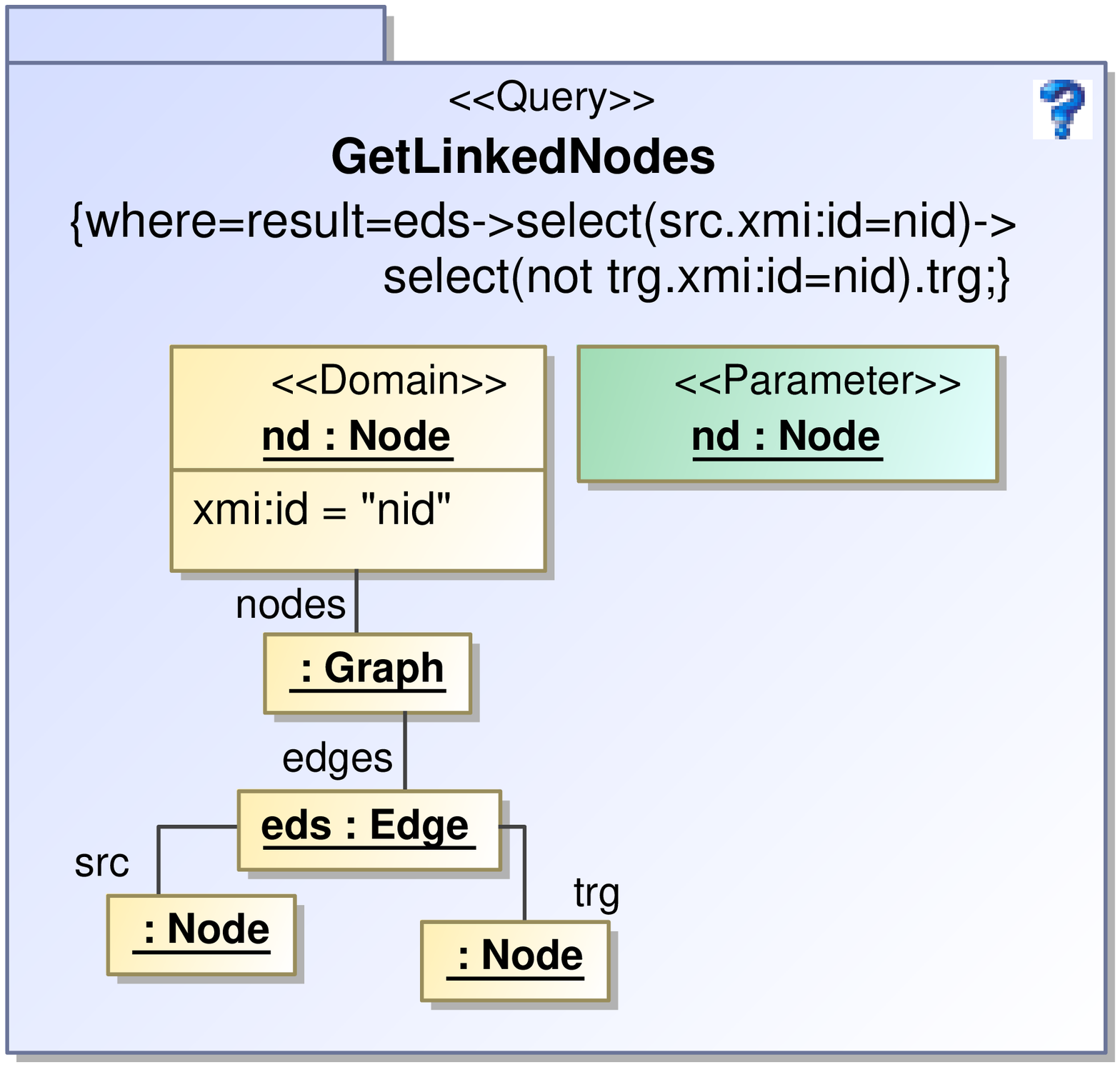}
\end{minipage}%
\begin{minipage}[t]{0.5\linewidth}
\vspace{0pt} \centering
\includegraphics[width=.9\linewidth]{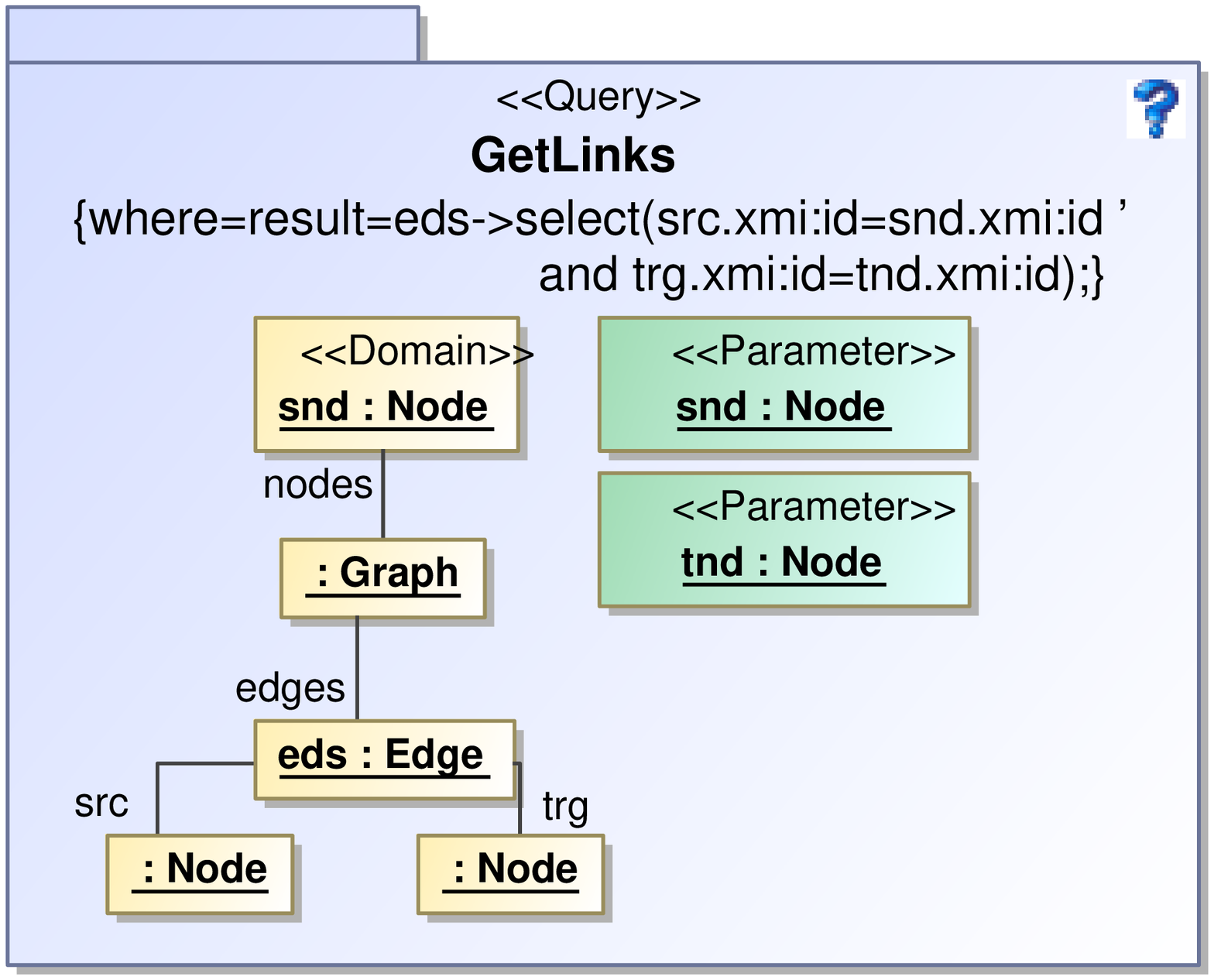}
\end{minipage}%
\\
\vspace*{-1.5\baselineskip}
\begin{minipage}[c]{0.5\linewidth}
\caption{Get target nodes of a node}
\label{fig:GetLinkedNodes_It}
\end{minipage}%
\begin{minipage}[c]{0.5\linewidth}
   \caption{Get edges between nodes}
\label{fig:GetLinks_It}
\end{minipage}
\end{figure}

\hide{
===================================
}